\documentclass{ieeeaccess}
\usepackage{cite}
\usepackage{amsmath,amssymb,amsfonts}
\usepackage{siunitx}
\usepackage{multicol}
\usepackage{multirow}
\usepackage{subcaption}
\usepackage{algorithmic}
\usepackage{graphicx}
\usepackage[breaklinks]{hyperref}
\usepackage{diagbox}
\def\BibTeX{{\rm B\kern-.05em{\sc i\kern-.025em b}\kern-.08em
    T\kern-.1667em\lower.7ex\hbox{E}\kern-.125emX}}
\begin{document}
\history{Date of latest submission January 30, 2021}
\doi{}
\title{A Hitchhiker's Guide to Structural Similarity}

\author{\uppercase{Abhinau K. Venkataramanan\authorrefmark{1}, Chengyang Wu\authorrefmark{1}, Alan C. Bovik\authorrefmark{1}\IEEEmembership{Fellow, IEEE}, Ioannis Katsavounidis\authorrefmark{2}, and Zafar Shahid\authorrefmark{2}}}

\address[1]{Department of Electrical and Computer Engineering, University of Texas at Austin, Austin, TX 78705 USA}
\address[2]{Facebook, Menlo Park, CA 94025 USA}
\tfootnote{This research was supported by funding from Facebook Video Infrastructure}

\markboth
{Venkataramanan \headeretal: A Hitchhiker's Guide to Structural Simlarity}
{Venkataramanan \headeretal: A Hitchhiker's Guide to Structural Simlarity}

\corresp{Corresponding author: Abhinau K. Venkataramanan (e-mail: abhinaukumar@utexas.edu).}

\begin{abstract}
The Structural Similarity (SSIM) Index is a very widely used image/video quality model that continues to play an important role in the perceptual evaluation of compression algorithms, encoding recipes and numerous other image/video processing algorithms. Several public implementations of the SSIM and Multiscale-SSIM (MS-SSIM) algorithms have been developed, which differ in efficiency and performance. This ``bendable ruler" makes the process of quality assessment of encoding algorithms unreliable. To address this situation, we studied and compared the functions and performances of popular and widely used implementations of SSIM, and we also considered a variety of design choices. Based on our studies and experiments, we have arrived at a collection of recommendations on how to use SSIM most effectively, including ways to reduce its computational burden.
\end{abstract}


\titlepgskip=-15pt

\maketitle

\section{Introduction}
\label{sec:introduction}
With the explosion of social media platforms and online streaming services, video has become the most widely consumed form of content on the internet, accounting for 60\% of global internet traffic in 2019 \cite{ref:video_internet_report}. Social media platforms have also led to an explosion in the amount of image data being shared and stored online. Handling such large volumes of image and video data is inconceivable without the use of compression algorithms such as JPEG \cite{ref:jpeg} \cite{ref:jpeg_book}, AVIF (AV1 Intra) \cite{ref:avif_netflix} \cite{ref:avif_eval}, HEIF (HEVC Intra) \cite{ref:heif}, H.264 \cite{ref:h264} \cite{ref:h264_book}, HEVC \cite{ref:hevc}, EVC \cite{ref:evc}, VP9 \cite{ref:vp9}, AV1 \cite{ref:av1}, SVT-AV1 \cite{ref:svt_av1}, and the upcoming VVC and AV2 standards.

The goal of these algorithms is to perform lossy compression of images and videos to significantly reduce file sizes and bandwidth consumption, while incurring little or acceptable reduction of visual quality. In addition to compression-distorted streaming videos, a large fraction of the images and videos that are shared on social media are User Generated Content (UGC) \cite{ref:wild_db} \cite{ref:ugc_db}, i.e., not professionally created. As a result, even without any additional processing, these images and videos can have impaired quality because they were captured by uncertain hands. In all these circumstances, it is imperative to have available automatic perceptual quality models and algorithms which can accurately, reliably, and consistently predict the subjective quality of images/videos over this wide range of applications.

One way that perceptual quality models can provide significant gains in compression is by conducting perceptual Rate-Distortion Optimization RDO \cite{ref:dyn_optimizer}, where quantization parameters, encoding ``recipes," and mode decisions are evaluated by balancing the resulting bitrates against the perceptual quality of the decoded videos. Typically, a set of viable encodes is arrived at by constructing a perceptually-guided, Pareto-optimal bitrate ladder.

To understand Pareto-optimality, consider two encodes of a video \(E_1 = (R_1, D_1)\) and \(E_2 = (R_2, D_2)\), where \(R\) and \(D\) denote the bitrate and the (perceptual) distortion associated with each encode. If \(R_1 \leq R_2\) and \(D_1 \leq D_2\), we say that \(E_1\) ``dominates" \(E_2\), since better performance (lower distortion) is obtained at a lower cost (bitrate). So, we can prune any set of encodes \(S = \{E_i\}\), by removing all those dominated by any other encode. The pruned set, say \(S'\), has the property that for any two encodes \(E_1\) and \(E_2\) such that \(R_1 < R_2\), \(D_1 > D_2\). That is, we obtain a set of encodes such that an increase in bitrate corresponds to a decrease in distortion. Such a set is said to be Pareto-optimal. In general, we can define Pareto-optimality for any setting where a cost is incurred (bitrate, running time, etc.) to achieve better performance (accuracy, distortion, etc.)

The first distortion/quality metric used to measure image/video quality was the humble Peak Signal to Noise Ratio (PSNR), or log-reciprocal Mean Squared Error (MSE) between a reference video and possibly distorted versions of it (e.g., by compression). However, while the PSNR metric is simple and easy to calculate, it does not correlate very well with subjective opinion scores of picture or video quality \cite{ref:psnr_corr}. This is because PSNR is a pixel-wise fidelity metric which does not account for spatial or temporal perceptual processes.

An important breakthrough on the perceptual quality problem emerged in the form of the Universal Quality Index (UQI) \cite{ref:uqi}, the first form of the Structural Similarity Index (SSIM). Given a pair of images (reference and distorted), UQI creates a local quality map, by measuring luminance, contrast and structural similarity over local neighborhoods, then pooling (averaging) values of this spatial quality map yielding a single quality prediction (per picture or video frame). SSIM was later refined to better account for the interplay between adaptive gain control of the visual signal (the basis for masking effects) and saturation at low signal levels \cite{ref:ssim}. 

The SSIM concept reached a higher performance in the form of Multi-Scale SSIM (MS-SSIM) \cite{ref:msssim}, which applies SSIM at five spatial resolutions obtained by successive dyadic sampling. Contrast and structure similarities are computed at each scale, while luminance similarity is only calculated at the coarsest scale. These scores are then combined using exponential weighting. SSIM and MS-SSIM are widely used by the streaming and social media industries to perceptually control the encodes of many billions of picture and video contents annually.

While SSIM and MS-SSIM are most commonly deployed on a frame-by-frame basis, temporal extensions of SSIM have also been developed. In \cite{ref:ssim_video}, the authors compute frame-wise quality scores, weighted by the amount of motion in each frame. In \cite{ref:motion_fields_ssim}, explicit motion fields are used to compute SSIM along motion trajectories, an idea that was elaborated on in the successful MOVIE index \cite{ref:movie}.

Following the success of SSIM, a great variety of picture and video quality models have been proposed. Among these, the most successful have relied on perceptually relevant ``natural scene statistics" (NSS) models, which accurately and reliably describe bandpass and nonlinearly normalized visual signals \cite{ref:gsm_nss} \cite{ref:rr_nr_qa}. Distortions predictably alter these statistics, making it possible to create highly competitive picture and video quality predictors like the Full-Reference (FR) Visual Information Fidelity (VIF) index \cite{ref:vif}, the Spatio-Temporal Reduced Reference Entropic Differences (ST-RRED) model \cite{ref:strred}, the efficient SpEED-QA \cite{ref:speedqa}, and the Video Multi-method Assessment Fusion (VMAF) \cite{ref:vmaf} model, which uses a simple learning model to fuse quality features derived from NSS models to obtain high performance and widespread industry adoption. Despite these advances, SSIM remains the most widely used perceptual quality algorithm because of its high performance, natural definition, and compute simplicity. Moreover, the success of SSIM can also be explained by NSS, at least in part \cite{ref:unifying}.

In many situations, reference information is not available as a ``gold standard" against which the quality of a test picture or video can be evaluated. No-Reference (NR) quality models have been developed that can accurately predict picture or video quality without a reference, by measuring NSS deviations. Notable NR quality models include BLIINDS \cite{ref:bliinds}, DIIVINE \cite{ref:diivine}, BRISQUE \cite{ref:brisque}, and NIQE \cite{ref:niqe}. The latter two, which have attained significant industry penetration, are similar to SSIM since they are defined by simple bandpass operations over multiple scales, normalized by local spatial energy. For encoding applications where the source video to be encoded is already impaired by some distortion(s), e.g. UGC, as is often found on sites like YouTube, Facebook, and Instagram, SSIM and NIQE can be combined via a 2-step assessment process to produce significantly improved encode quality predictions \cite{ref:2step_patent} \cite{ref:2stepqa}.

Evaluating picture and video encodes at scale remains the most high-volume application of quality assessment, and SSIM continues to play a dominant role in this space. Nevertheless, many widely used versions of SSIM exist having different characteristics. Understanding and unifying these various implementations would be greatly useful to industry. Moreover, there remain questions regarding the use of SSIM across different display sizes, devices and viewing distances, as well as how to handle color, and how to combine (pool) SSIM scores. Our objective here is to attempt to answer these questions, at least in part.

\section{Background}
\label{sec:background}
The basic SSIM index is a FR picture quality model defined between two luminance images of size \(M \times N\), \(I_1(i,j)\) and \(I_2(i,j)\) as a multiplicative combination of three terms - luminance similarity \(l(i,j)\), contrast similarity \(c(i,j)\) and structure similarity \(s(i,j)\). Color may be considered, but we will do so later.

These three terms are defined in terms of the local means \(\mu_1(i,j), \mu_2(i,j)\), standard deviations \(\sigma_1(i,j), \sigma_2(i,j)\), and correlations \(\sigma_{12}(i,j)\) of luminance, as follows. Let \(\mathcal{W}_{ij}\) denote a windowed region of size \(k \times k\) spanning the indices \(\{i,\dots,i+k-1\} \times \{j,\dots,j+k-1\}\) and let \(w(m,n)\) denote weights assigned to each index \((m,n)\) of this window. In practice, these weighting functions sum to unity, and have a finite-extent Gaussian or rectangular shape.

The local statistics are then calculated on (and between) the two images as
\begin{equation}
    \mu_1(i,j) = \sum\limits_{m,n \in W_ij} w(m,n) I_1(m,n),
    \label{eq:local_means}
\end{equation}
\begin{equation}
    \mu_2(i,j) = \sum\limits_{m,n \in W_ij} w(m,n) I_2(m,n),
\end{equation}
\begin{equation}
    \sigma^2_1(i,j) = \sum\limits_{m,n \in W_ij} w(m,n) I_1^2(m,n) - \mu^2_1(i,j),
\end{equation}
\begin{equation}
    \sigma^2_2(i,j) = \sum\limits_{m,n \in W_ij} w(m,n) I_2^2(m,n) - \mu^2_2(i,j),
\end{equation}
\begin{equation}
    \sigma_{12}(i,j) = \sum\limits_{m,n \in W_ij} w(m,n) I_1(m,n)I_2(m,n) - \mu_1(i,j)\mu_2(i,j).
    \label{eq:local_corr}
\end{equation}

Using these local statistics, the luminance, contrast and structural similarity terms are respectively defined as
\begin{equation}
    l(i,j) = \frac{2\mu_1(i,j)\mu_2(i,j) + C_1}{\mu^2_1(i,j) + \mu^2_2(i,j) + C_1},
    \label{eq:lum_term}
\end{equation}
\begin{equation}
    c(i,j) = \frac{2\sigma_1(i,j)\sigma_2(i,j) + C_2}{\sigma^2_1(i,j) + \sigma^2_2(i,j) + C_2},
    \label{eq:cont_term}
\end{equation}
\begin{equation}
    s(i,j) = \frac{\sigma_{12}(i,j) + C_3}{\sigma_1(i,j)\sigma_2(i,j) + C_3},
\end{equation}
where \(C_1, C_2\) and \(C_3\) are saturation constants that contribute to numerical stability. Local quality scores are then defined as
\begin{equation}
    Q(i,j) = l(i,j) \cdot c(i,j) \cdot s(i,j).
\end{equation}

Adopting the common choice of \(C_3 = C_2/2\), the contrast and structure terms combine:
\begin{equation}
    cs(i,j) = c(i,j) s(i,j) = \frac{2\sigma_{12}(i,j) + C_2}{\sigma^2_1(i,j) + \sigma^2_2(i,j) + C_2}.
\end{equation}

In this way, a SSIM quality map \(Q(i,j)\) is defined, which can be used to visually localize distortions. Since a single picture quality score is usually desired, the average value of the quality map can be reported as the Mean SSIM (MSSIM) score between the two images:
\begin{equation}
    SSIM(I_1,I_2) = \frac{1}{MN} \sum_{i=1}^{M}\sum_{j=1}^{N} Q(i,j).
\end{equation}

SSIM obeys the following desirable properties:
\begin{enumerate}
    \item Symmetry: \(SSIM(I_1,I_2) = SSIM(I_2,I_1)\)
    \item Boundedness: \(|SSIM(I_1,I_2)| \leq 1\footnote{Very rarely, distortion can cause a negative correlation between reference and test image patches.}\)
    \item Unique Maximum: \(SSIM(I_1,I_2) = 1 \iff I_1 = I_2\)
\end{enumerate}

An important property of SSIM is that it accounts for the perceptual phenomenon of Weber's Law, whereby a Just Noticeable Difference (JND) is proportional to the local neighborhood property \(Q\). This is the basis for perceptual masking of distortions, whereby the visibility of a distortion \(\Delta Q\) is mediated by the relative perturbation \(\Delta Q/Q\). 

To illustrate the connection between SSIM and Weber masking, consider an error \(\Delta \mu\) of local luminance \(\mu_1\) in a test image:
\begin{equation}
\mu_2 = \mu_1 + \Delta \mu = \mu_1(1 + \Lambda),
\end{equation}
where \(\Lambda = \Delta \mu / \mu_1\) is the relative change in luminance. Then, the luminance similarity term \eqref{eq:lum_term} becomes (dropping spatial indices)
\begin{align}
    l &= \frac{2\mu_1\mu_2 + C_1}{\mu_1^2 + \mu_2^2 + C_1} \\
    &= \frac{2\mu_1^2(1 + \Lambda) + C_1}{\mu_1^2\left(1 + (1 + \Lambda)^2\right) + C_1} \\
    &= \frac{2(1 + \Lambda) + C_1/\mu_1^2}{1 + (1 + \Lambda)^2 + C_1/\mu_1^2}.
\end{align}

Since it is usually true that \(C_1\ll\mu_1^2\), the luminance term \(l\) is approximately only a function of the relative luminance change, reflecting luminance masking.

Similarly, a locally perturbed contrast \(\sigma_1\) in a test image may be expressed as \(\sigma_2 = (1 + \Sigma) \sigma_1\), where \(\Sigma = \Delta \sigma / \sigma_1\) is the relative change in contrast from distortion. Similar to the above, we can express the contrast term \eqref{eq:cont_term} as
\begin{equation}
    c = \frac{2(1 + \Sigma) + C_2/\sigma_1^2}{1 + (1 + \Sigma)^2 + C_2/\sigma_1^2}.
\end{equation}
Since generally \(C_2 \ll \sigma_1^2\), the contrast term \(c\) is approximately a function of the relative, rather than absolute, change in contrast, thereby accounting the perceptual contrast masking.

Given an 8-bit luminance image, assume the nominal dynamic range \([0, L]\), where \(L = 255\). Most commonly, the saturation constants are chosen relative to the dynamic range as \(C_1 = (K_1 L)^2\) and \( C_2 = (K_2  L)^2\), where \(K_1\) and \(K_2\) are small constants.

SSIM is quite flexible and allows room for design choices. The recommended implementation of SSIM in \cite{ref:ssim} is
\begin{itemize}
    \item If \(\min(M,N)>256\), resize images such that \(\min(M,N) = 256\).
    \item Use a Gaussian weighting window in \eqref{eq:local_means} - \eqref{eq:local_corr} having \(k=11\) and \(\sigma=1.5\).
    \item Choose regularization constants \(K_1 = 0.01, K_2=0.03\).
\end{itemize}

One of our goals here is to  compare and test different, commonly used implementations of SSIM and MS-SSIM, which make different design choices. We conduct performance evaluations on existing, well-regarded image and video quality databases. We study the effects of several design choices and make recommendations on best practices when utilizing SSIM.

\section{Databases}
\label{sec:databases}
One of our main goals is to help ``standardize" the way SSIM is defined and used. Since many versions of SSIM exist and implementing SSIM involves design choices, reliable and accurate subjective test beds that capture the breadth of theoretical and practical distortions are the indispensable tools for our analysis. Among these, we selected two picture quality databases and two video quality databases that are widely used.

\subsection{LIVE Image Quality Assessment Database}
The LIVE IQA database \cite{ref:live_iqa_db_link} \cite{ref:live_iqa_db_pap1} contains 29 reference pictures, each subjected to the following five distortions (each at four levels of severity).
\begin{itemize}
    \item JPEG compression
    \item JPEG2000 compression
    \item Gaussian blur
    \item White noise
    \item Bit errors in JPEG2000 bit streams
\end{itemize}
LIVE IQA contains 982 pictures with nearly 30,000 corresponding Difference Mean Opinion (DMOS) human subject scores.

\subsection{Tampere Image Database 2013}
The Tampere Image Database 2013 (TID2013) \cite{ref:tid2013_db} contains 3000 distorted pictures subjected to 24 impairments at 5 distortion levels synthetically applied to 25 pristine images. The 24 distortions are
\begin{itemize}
    \item Additive Gaussian noise
    \item Additive noise in color components is more intensive than additive noise in the luminance component
    \item Spatially correlated noise
    \item Masked noise
    \item High frequency noise
    \item Impulse noise
    \item Quantization noise
    \item Gaussian blur
    \item Image denoising
    \item JPEG compression
    \item JPEG2000 compression
    \item JPEG transmission errors
    \item JPEG2000 transmission errors
    \item Non eccentricity pattern noise
    \item Local block-wise distortions of different intensity
    \item Mean shift (intensity shift)
    \item Contrast change
    \item Change of color saturation
    \item Multiplicative Gaussian noise
    \item Comfort noise
    \item Lossy compression of noisy images
    \item Image color quantization with dither
    \item Chromatic aberrations
    \item Sparse sampling and reconstruction
\end{itemize}
The 3000 pictures in TID 2013 are accompanied by human subjective quality scores on them in the form of over 500,000 MOS.

\subsection{LIVE Video Quality Assessment Database}
The LIVE VQA database \cite{ref:live_vqa_db} contains 10 reference videos, each subjected to the following four distortions (each applied at four levels of severity).
\begin{itemize}
    \item MPEG-2 compression
    \item H.264 compression
    \item Error-prone IP networks
    \item Error-prone wireless networks
\end{itemize}
A total of 150 distorted videos are obtained, on which 4350 subjective DMOS were obtained.

\subsection{Netflix Public Database}
The Netflix Public Database, obtained from the VMAF \cite{ref:vmaf} Github repository, contains 9 reference videos, each distorted by spatial scaling and compression, yielding 70 distorted videos. We selected this database because of its high relevance and commonly observed distortions characteristic of SSIM streaming video deployments at the largest scales.

\begin{table*}[ht]
\caption{Salient features of SSIM implementations}
\label{tab:ssim_implementation_details}
\centering
\begin{tabular}{|c|c|c|c|c|}
    \hline
    Implementation & Window Function & Scaling & Border Handling & Comments \\
    \hline
    FFMPEG & \begin{minipage}[t]{0.45\columnwidth}{8x8 rectangular window, with a stride of 4.}\end{minipage} & \begin{minipage}[t]{0.4\columnwidth}{None.}\end{minipage} & \begin{minipage}[t]{0.3\columnwidth}{Only valid convolution outputs are used.}\end{minipage} & \begin{minipage}[t]{0.4\columnwidth}{None.}\end{minipage} \\
    \hline
    LIBVMAF & \begin{minipage}[t]{0.45\columnwidth}{11x11 Gaussian window}\end{minipage} & \begin{minipage}[t]{0.4\columnwidth}{Downsampled by a factor of \(\max(1, [\min(W,H)/256)])\).}\end{minipage} & \begin{minipage}[t]{0.3\columnwidth}{Only valid convolution outputs are used.}\end{minipage} & \begin{minipage}[t]{0.4\columnwidth}{Only luminance channel is processed.}\end{minipage} \\
    \hline
    ClearView & \begin{minipage}[t]{0.45\columnwidth}{Gaussian window}\end{minipage} & \begin{minipage}[t]{0.4\columnwidth}{Unknown.}\end{minipage} & \begin{minipage}[t]{0.3\columnwidth}{Zero padding is used to extend borders.}\end{minipage} & \begin{minipage}[t]{0.4\columnwidth}{Images of arbitrary resolutions are not supported. A configurable threshold is used to filter frame scores.}\end{minipage} \\
    \hline
    HDRTools & \begin{minipage}[t]{0.45\columnwidth}{Strided Rectangular window (8x8 with stride 1, by default).}\end{minipage} & \begin{minipage}[t]{0.4\columnwidth}{None.}\end{minipage} & \begin{minipage}[t]{0.3\columnwidth}{Only valid convolution outputs are used.}\end{minipage} & \begin{minipage}[t]{0.4\columnwidth}{None.}\end{minipage} \\
    \hline
    Daala &  \begin{minipage}[t]{0.45\columnwidth}{Gaussian window truncated to a value of 0.5, i.e., window size \(k = \) \\ \(2 \sigma\left(-2\log \left(\sigma\sqrt{\frac{\pi}{2}}\right)\right)^{1/2}+1\).}\end{minipage} & \begin{minipage}[t]{0.4\columnwidth}{Using a Gaussian window having standard deviation \(\sigma = 1.5h/256\).}\end{minipage} & \begin{minipage}[t]{0.3\columnwidth}{Only valid convolution outputs are used.}\end{minipage} & \begin{minipage}[t]{0.4\columnwidth}{None.}\end{minipage} \\
    \hline
    Scikit-Image & \begin{minipage}[t]{0.45\columnwidth}{Rectangular or Gaussian window (11x11, by default).}\end{minipage} & \begin{minipage}[t]{0.4\columnwidth}{None.}\end{minipage} & \begin{minipage}[t]{0.3\columnwidth}{Reflection padding is used to extend borders.}\end{minipage} & \begin{minipage}[t]{0.4\columnwidth}{None.}\end{minipage} \\
    \hline
    Scikit-Video & \begin{minipage}[t]{0.45\columnwidth}{Any window (user-input).}\end{minipage} & \begin{minipage}[t]{0.4\columnwidth}{Downsampled by a factor of \(\max(1, [\min(W,H)/256)])\).}\end{minipage} & \begin{minipage}[t]{0.3\columnwidth}{Zero padding is used to extend borders.}\end{minipage} & \begin{minipage}[t]{0.4\columnwidth}{Irrespective of window size, 5 border pixels are removed.}\end{minipage} \\
    \hline
    Tensorflow & \begin{minipage}[t]{0.45\columnwidth}{Gaussian window.}\end{minipage} & \begin{minipage}[t]{0.4\columnwidth}{None.}\end{minipage} & \begin{minipage}[t]{0.3\columnwidth}{Only valid convolution outputs are used.}\end{minipage} & \begin{minipage}[t]{0.4\columnwidth}{None.}\end{minipage} \\
    \hline
    FastQA & \begin{minipage}[t]{0.45\columnwidth}{Strided rectangular window (11x11 with stride 1, by default).}\end{minipage} & \begin{minipage}[t]{0.4\columnwidth}{None.}\end{minipage} & \begin{minipage}[t]{0.3\columnwidth}{Only valid convolution outputs are used.}\end{minipage} & \begin{minipage}[t]{0.4\columnwidth}{A custom implementation we created that uses integral images. Publicly available \href{https://github.com/abhinaukumar/fast-qa}{here}.}\end{minipage} \\
    \hline
    MATLAB & \begin{minipage}[t]{0.45\columnwidth}{Gaussian window. Size is inferred from standard deviation \(\sigma\) as \(2\times\lceil3\sigma\rceil+1\).}\end{minipage} & \begin{minipage}[t]{0.4\columnwidth}{None.}\end{minipage} & \begin{minipage}[t]{0.3\columnwidth}{Only valid convolution outputs are used.}\end{minipage} & \begin{minipage}[t]{0.4\columnwidth}{None.}\end{minipage} \\
    \hline
    MATLAB (Fast) & \begin{minipage}[t]{0.45\columnwidth}{Rectangular window.}\end{minipage} & \begin{minipage}[t]{0.4\columnwidth}{None.}\end{minipage} & \begin{minipage}[t]{0.3\columnwidth}{Only valid convolution outputs are used.}\end{minipage} & \begin{minipage}[t]{0.4\columnwidth}{Integer arithmetic is used to accelerate computations.}\end{minipage} \\
    \hline
\end{tabular}
\end{table*}

\section{Versions of SSIM}
\label{sec:ssim_versions}
Next, we take a deep dive into publicly available and commonly used implementations of SSIM and MS-SSIM. We compare various aspects of their performance, explain the differences between them, and provide recommendations for best practices when using SSIM. This is especially important because, as we will see, subtle differences in design choices can lead to significant changes in performance and efficiency.

\subsection{Public SSIM and MS-SSIM Implementations}
We considered the following ten SSIM implementations when carrying out our experiments:
\begin{enumerate}
    \item FFMPEG \cite{ref:ffmpeg}
    \item LIBVMAF \cite{ref:vmaf}
    \item VideoClarity ClearView Player (ClearView)
    \item HDRTools \cite{ref:hdrtools}
    \item Daala \cite{ref:daala}
    \item Scikit-Image in Python (Rectangular) \cite{ref:skimage}
    \item Scikit-Image in Python (Gaussian) \cite{ref:skimage}
    \item Scikit-Video in Python (Rectangular) \cite{ref:skvideo}
    \item Scikit-Video in Python (Gaussian) \cite{ref:skvideo}
    \item Tensorflow in Python \cite{ref:tf}
    \item MATLAB
    \item \label{item:matlab_fast} MATLAB (Fast)
\end{enumerate}

``Rectangular" refers to using a constant weight window function to calculate local statistics, while ``Gaussian" refers to using a Gaussian-shaped window function to calculate local statistics, as in \cite{ref:ssim}. Only the Python and MATLAB implementations allow the user to set parameters such as the SSIM window size. Hence, we tested the other implementations using the default parameters. ``Fast'' in item~\ref{item:matlab_fast} refers to an accelerated implementation of SSIM in MATLAB.

In addition, the following eight MS-SSIM implementations were tested:
\begin{enumerate}
    \item LIBVMAF
    \item ClearView
    \item HDRTools
    \item Daala
    \item \label{item:daala_fast} Daala (Fast)
    \item \label{item:skvideo_sum_multiscale} Scikit-Video in Python (Sum)
    \item \label{item:skvideo_product_multiscale} Scikit-Video in Python (Product)
    \item Tensorflow in Python
\end{enumerate}

``Sum" and ``Product" in~\ref{item:skvideo_sum_multiscale} and~\ref{item:skvideo_product_multiscale} refer to different ways of aggregating SSIM scores across scales. ``Product" refers to the method proposed in \cite{ref:msssim}, where MS-SSIM is computed as an exponentially-weighted product of SSIM scores from each scale. In ``Sum", the MS-SSIM score is instead a weighted average of SSIM scores across scales. ``Fast'' refers to an accelerated implementation of MS-SSIM in Daala.

\subsection{Salient Features of SSIM and MS-SSIM Implementations}

\begin{table*}[ht]
\caption{Off-the-shelf performance of SSIM implementations}
\label{tab:default_params}
\begin{subtable}{\linewidth}
    \centering
    \caption{Performance of SSIM implementations}
    \label{tab:ssim_default_perfs}
    \begin{tabular}{|c|c|c|c|c|c|c|c|c|c|}
    \hline
    \multirow{2}{*}{Implementation} & \multicolumn{3}{c|}{LIVE IQA} & \multicolumn{3}{c|}{TID 2013} & \multicolumn{3}{c|}{LIVE VQA}\\
    \cline{2-10}
    & PCC & SROCC & 1-RMSE & PCC & SROCC & 1-RMSE & PCC & SROCC & 1-RMSE \\
    \hline
    FFMPEG & \textbf{0.942} & 0.931 & 0.908 & 0.707 & 0.658 & 0.874 & 0.603 & 0.599 & 0.826 \\
    \hline
    LIBVMAF & \textbf{0.946} & \textbf{0.946} & \textbf{0.912} & \textbf{0.802} & \textbf{0.755} & \textbf{0.894} & \textbf{0.700} & \textbf{0.695} & \textbf{0.844} \\
    \hline
    Daala SSIM & 0.940 & 0.929 & 0.907 & 0.701 & 0.657 & 0.873 & 0.621 & 0.618 & 0.829 \\
    \hline
    ClearView & 0.791 & 0.789 & 0.883 & 0.718 & 0.683 & 0.876 & 0.455 & 0.376 & 0.805 \\
    \hline
    HDRTools & 0.845 & 0.831 & 0.898 & 0.667 & 0.605 & 0.868 & 0.471 & 0.452 & 0.807 \\
    \hline
    Scikit-Image (Rect) & 0.942 & 0.930 & 0.908 & 0.692 & 0.639 & 0.872 & 0.665 & 0.668 & 0.837 \\
    \hline
    Scikit-Image (Gauss) & 0.939 & 0.925 & 0.906 & 0.677 & 0.628 & 0.869 & 0.563 & 0.549 & 0.819 \\
    \hline
    Scikit-Video (Rect) & 0.944 & \textbf{0.950} & 0.910 & \textbf{0.793}& \textbf{0.744} & \textbf{0.892} & \textbf{0.753} & \textbf{0.745} & \textbf{0.856} \\
    \hline
    Scikit-Video (Gauss) & 0.945 & \textbf{0.950} & \textbf{0.911} & \textbf{0.790} & \textbf{0.742} & \textbf{0.891} & \textbf{0.710} & \textbf{0.703} & \textbf{0.846} \\
    \hline
    FastQA & \textbf{0.946} & 0.934 & \textbf{0.911} & 0.711 & 0.661 & 0.875 & 0.656 & 0.661 & 0.835 \\
    \hline
    MATLAB (Fast) & 0.864 & 0.851 & 0.904 & 0.687 & 0.627 & 0.871 & 0.577 & 0.531 & 0.821 \\
    \hline
    MATLAB & 0.862 & 0.848 & 0.903 & 0.686 & 0.624 & 0.871 & 0.563 & 0.521 & 0.819 \\
    \hline
    Tensorflow & 0.939 & 0.925 & 0.906 & 0.677 & 0.628 & 0.869 & 0.562 & 0.549 & 0.819 \\
    \hline
    \end{tabular}
\end{subtable}
\\
\linebreak
\\
\begin{subtable}{\linewidth}
    \centering
    \caption{Performance of MS-SSIM Implementations}
    \label{tab:ssim_multiscale_default_perfs}
    \begin{tabular}{|c|c|c|c|c|c|c|c|c|c|}
    \hline
    \multirow{2}{*}{Implementation} & \multicolumn{3}{c|}{LIVE IQA} & \multicolumn{3}{c|}{TID 2013} & \multicolumn{3}{c|}{LIVE VQA} \\
    \cline{2-10}
    & PCC & SROCC & 1-RMSE & PCC & SROCC & 1-RMSE & PCC & SROCC & 1-RMSE \\
    \hline
    LIBVMAF & 0.943 & 0.946 & 0.909 & \textbf{0.837} & \textbf{0.788} & \textbf{0.903} & 0.737 & 0.732 & 0.852 \\
    \hline
    Daala & \textbf{0.946} & 0.949 & 0.911 & 0.832 & \textbf{0.792} & 0.901 & 0.739 & 0.734 & 0.853 \\
    \hline
    Daala Fast-SSIM & 0.942 & 0.940 & 0.909 & 0.782 & 0.728 & 0.889 & 0.654 & 0.631 & 0.835 \\
    \hline
    ClearView & 0.871 & 0.870 & 0.906 & 0.749 & 0.699 & 0.882 & 0.654 & 0.627 & 0.835 \\
    \hline
    HDRTools & 0.904 & 0.906 & \textbf{0.919} & \textbf{0.834} & \textbf{0.789} & \textbf{0.902} & 0.733 & 0.726 & 0.851 \\
    \hline
    Scikit-Video (Product) & 0.944 & \textbf{0.954} & 0.910 & 0.813 & 0.763 & 0.896 & \textbf{0.756} & \textbf{0.748} & \textbf{0.857} \\
    \hline
    Scikit-Video (Sum) & 0.943 & \textbf{0.954} & 0.909 & 0.810 & 0.762 & 0.896 & \textbf{0.760} & \textbf{0.748} & \textbf{0.858} \\
    \hline
    FastQA & \textbf{0.947} & \textbf{0.951} & \textbf{0.912} & 0.825 & 0.778 & 0.899 & \textbf{0.760} & \textbf{0.751} & \textbf{0.858} \\
    \hline
    MATLAB & 0.902 & 0.905 & 0.918 & 0.829 & 0.782 & 0.901 & 0.737 & 0.729 & 0.852 \\
    \hline
    Tensorflow & \textbf{0.947} & 0.951 & 0.912 & \textbf{0.833} & 0.786 & \textbf{0.902} & 0.745 & 0.743 & 0.854 \\
    \hline
    \end{tabular}
\end{subtable}
\end{table*}

The salient features of various SSIM implementations have been listed in Table~\ref{tab:ssim_implementation_details}, and the salient features of various MS-SSIM implementations have been listed below. To avoid repetition, we only discuss the aspects in which each MS-SSIM implementation deviates from the corresponding SSIM implementation.
\begin{enumerate}
    \item LIBVMAF
        \begin{enumerate}
            \item Dyadic down-sampling is performed using a 9/7 biorthogonal wavelet filter.
        \end{enumerate}
    \item Daala
        \begin{enumerate}
            \item Uses \(\sigma = 1.5\), leading to a Gaussian window of size 11.
            \item Dyadic down-sampling is performed by 2x2 average pooling.
        \end{enumerate}
    \item Daala (Fast)
        \begin{enumerate}
            \item Multiscale processing is performed at 4 levels. The first four exponents used in the standard MS-SSIM formulation are renormalized to sum to 1.
            \item An integer approximation to Gaussian is used.
            \item Dyadic down-sampling is performed by 2x2 average pooling.
            \item When image dimensions were not multiples of 16, we found that this implementation suffered from memory leaks, which led to a considerable decrease in accuracy. A simple fix restored performance to expected levels.
        \end{enumerate}
    \item Scikit-Video
        \begin{enumerate}
            \item Dyadic down-sampling is performed by low pass filtering using a 2x2 average filter, followed by down-sampling.
            \item Allows aggregating across scales by summation instead of product. For summation, the exponents \(\beta_i\) are normalized to sum to 1, leading to a convex combination.
            \item At the coarsest scale, the algorithm uses \(\alpha_M = \beta_M = \gamma_M = 1\).
            \item When image dimensions were large, we found that this implementation suffered from incompatible memory allocation, leading to crashes at run-time. We fixed this error, making the implementation usable.
        \end{enumerate}
    \item Tensorflow
        \begin{enumerate}
            \item Dyadic down-sampling is carried out by average pooling 2x2 neighborhoods.
        \end{enumerate}
\end{enumerate}

\subsection{Off-the-shelf Performance using Default Parameters}

\begin{table*}[ht]
\caption{Off-the-shelf performance of SSIM and MS-SSIM implementations on compression}
\label{tab:comp_default_params}
\begin{subtable}{\linewidth}
    \centering
    \caption{Performance of  SSIM Implementations}
    \label{tab:comp_ssim_default_perfs}
    \begin{tabular}{|c|c|c|c|c|c|c|c|c|c|c|c|c|}
    \hline
    \multirow{2}{*}{Implementation} & \multicolumn{3}{c|}{LIVE IQA (Comp)} & \multicolumn{3}{c|}{TID 2013 (Comp)} & \multicolumn{3}{c|}{LIVE VQA (Comp)} & \multicolumn{3}{c|}{Netflix Public}\\
    \cline{2-13}
    & PCC & SROCC & 1-RMSE & PCC & SROCC & 1-RMSE & PCC & SROCC & 1-RMSE & PCC & SROCC & 1-RMSE\\
    \hline
    FFMPEG & 0.970 & 0.967 & 0.930 & 0.938 & 0.926 & 0.918 & 0.652 & 0.652 & 0.845 & 0.696 & 0.657 & 0.797 \\
    \hline
    LIBVMAF & 0.941 & 0.954 & 0.902 & \textbf{0.961} & \textbf{0.947} & \textbf{0.935} & \textbf{0.689} & \textbf{0.684} & \textbf{0.852} & \textbf{0.768} & \textbf{0.765} & \textbf{0.819} \\
    \hline
    Daala SSIM & \textbf{0.970} & 0.968 & 0.930 & 0.938 & 0.926 & 0.919 & 0.657 & 0.658 & 0.846 & \textbf{0.707} & \textbf{0.680} & \textbf{0.800} \\
    \hline
    ClearView & 0.856 & 0.851 & 0.901 & 0.863 & 0.854 & 0.881 & 0.532 & 0.352 & 0.827 & 0.590 & 0.538 & 0.772 \\
    \hline
    HDRTools & 0.912 & 0.901 & 0.922 & 0.895 & 0.903 & 0.895 & 0.579 & 0.532 & 0.833 & 0.589 & 0.558 & 0.772 \\
    \hline
    Scikit-Image (Rect) & \textbf{0.971} & \textbf{0.969} & \textbf{0.932} & 0.944 & 0.928 & \textbf{0.923} & \textbf{0.686} & \textbf{0.691} & \textbf{0.851} & 0.681 & 0.656 & 0.793 \\
    \hline
    Scikit-Image (Gauss) & 0.969 & 0.965 & 0.929 & 0.924 & 0.917 & 0.910 & 0.637 & 0.622 & 0.842 & 0.642 & 0.621 & 0.783 \\
    \hline
    Scikit-Video (Rect) & 0.934 & 0.959 & 0.893 & 0.956 & 0.940 & 0.915 & 0.702 & 0.697 & 0.842 & 0.702 & 0.678 & 0.799 \\
    \hline
    Scikit-Video (Gauss) & 0.962 & \textbf{0.969} & 0.917 & 0.960 & 0.945 & 0.919 & 0.697 & 0.694 & 0.841 & 0.767 & 0.764 & 0.819 \\
    \hline
    FastQA & \textbf{0.972} & \textbf{0.969} & \textbf{0.930} & \textbf{0.946} & \textbf{0.930} & 0.906 & 0.670 & 0.680 & 0.836 & \textbf{0.787} & \textbf{0.778} & \textbf{0.826} \\
    \hline
    MATLAB (Fast) & 0.930 & 0.917 & \textbf{0.930} & 0.924 & 0.917 & 0.910 & 0.642 & 0.609 & 0.843 & 0.645 & 0.596 & 0.784 \\
    \hline
    MATLAB & 0.928 & 0.915 & 0.929 & 0.918 & 0.913 & 0.907 & 0.630 & 0.610 & 0.842 & 0.626 & 0.592 & 0.780 \\
    \hline
    Tensorflow & 0.969 & 0.965 & 0.929 & 0.924 & 0.917 & 0.910 & 0.636 & 0.624 & 0.842 & 0.642 & 0.621 & 0.783 \\
    \hline
    \end{tabular}
\end{subtable}
\\
\linebreak
\\
\begin{subtable}{\linewidth}
    \centering
    \caption{Performance of MS-SSIM Implementations}
    \label{tab:comp_ssim_multiscale_default_perfs}
    \begin{tabular}{|c|c|c|c|c|c|c|c|c|c|c|c|c|}
    \hline
    \multirow{2}{*}{Implementation} & \multicolumn{3}{c|}{LIVE IQA (Comp)} & \multicolumn{3}{c|}{TID 2013 (Comp)} & \multicolumn{3}{c|}{LIVE VQA (Comp)} & \multicolumn{3}{c|}{Netflix Public}\\
    \cline{2-13}
    & PCC & SROCC & 1-RMSE & PCC & SROCC & 1-RMSE & PCC & SROCC & 1-RMSE & PCC & SROCC & 1-RMSE\\
    \hline
    LIBVMAF & 0.931 & 0.956 & 0.895 & 0.967 & 0.943 & 0.940 & 0.693 & 0.694 & 0.852 & 0.754 & 0.739 & 0.814 \\
    \hline
    Daala & 0.936 & \textbf{0.962} & 0.898 & \textbf{0.968} & \textbf{0.952} & \textbf{0.941} & 0.692 & 0.696 & 0.852 & 0.755 & 0.741 & 0.815 \\
    \hline
    Daala Fast-SSIM & \textbf{0.949} & 0.947 & 0.909 & 0.926 & 0.900 & 0.911 & 0.595 & 0.565 & 0.835 & \textbf{0.769} & 0.749 & \textbf{0.819} \\
    \hline
    ClearView & 0.945 & 0.927 & \textbf{0.938} & 0.945 & 0.929 & 0.923 & \textbf{0.714} & \textbf{0.696} & \textbf{0.857} & \textbf{0.812} & \textbf{0.794} & \textbf{0.835} \\
    \hline
    HDRTools & \textbf{0.946} & 0.927 & \textbf{0.938} & \textbf{0.974} & \textbf{0.957} & \textbf{0.947} & \textbf{0.698} & 0.683 & \textbf{0.854} & 0.755 & 0.738 & 0.815 \\
    \hline
    Scikit-Video (Product) & 0.930 & 0.962 & 0.894 & \textbf{0.967} & \textbf{0.950} & \textbf{0.940} & 0.692 & 0.686 & 0.852 & 0.755 & 0.752 & 0.815 \\
    \hline
    Scikit-Video (Sum) & \textbf{0.961} & \textbf{0.967} & 0.920 & 0.967 & 0.949 & 0.940 & 0.697 & 0.688 & 0.853 & 0.754 & \textbf{0.751} & 0.814 \\
    \hline
    FastQA & 0.928 & 0.959 & 0.888 & 0.955 & 0.943 & 0.914 & 0.696 & \textbf{0.696} & 0.841 & \textbf{0.774} & \textbf{0.762} & \textbf{0.821} \\
    \hline
    MATLAB & 0.942 & 0.925 & \textbf{0.936} & 0.966 & 0.949 & 0.939 & \textbf{0.701} & 0.685 & \textbf{0.854} & 0.753 & 0.739 & 0.814 \\
    \hline
    Tensorflow & 0.942 & \textbf{0.962} & 0.904 & 0.966 & 0.949 & 0.939 & 0.698 & \textbf{0.701} & 0.853 & 0.754 & 0.746 & 0.814 \\
    \hline
    \end{tabular}
\end{subtable}
\end{table*}

In this section, we evaluate the off-the-shelf performance of the implementations discussed above. We first normalized the subjective scores of pictures/videos each database to the range \([0,1]\) by scaling and shifting. In all of the experiments in this section, unless mentioned otherwise, we computed SSIM scores only on the luminance channel. 

It is well known that the relationship between SSIM (or any other quality metric) and subjective scores is non-linear. To account for this, we fit the five-parameter logistic (5PL) function \cite{gottschalk2005five} shown in \eqref{eq:five_param_logistic} from SSIM values to subjective scores:
\begin{equation}
    Q(x) = \beta_1 \left( \frac{1}{2} - \frac{1}{1 + \exp (\beta_2(x - \beta_3))}\right) + \beta_4 x + \beta_5,
    \label{eq:five_param_logistic}
\end{equation}
where \(x\) is the SSIM score, \(\beta_i\) are parameters of the logistic function, and \(Q(x)\) is the predicted subjective quality.

After linearizing the SSIM values in this manner, we report the Pearson Correlation Coefficient (PCC) which is a measure of the linear correlation between the predicted and true quality, the Spearman Rank Order Correlation Coefficient (SROCC) which is a measure of the rank correlation (monotonicity), and the Root Mean Square Error (RMSE) which is a measure of the error in predicting subjective quality.

Table~\ref{tab:default_params} shows the results of the experiments, and the best three results in each column have been boldfaced. Among SSIM implementations, LIBVMAF and the Scikit-Video implementations generally outperformed all other algorithms. We attribute this superior performance to the use of scaling, which we will expound in Section~\ref{sec:device_dependency}.

Among the MS-SSIM implementations, there was no consistent "winners." Python implementations like Scikit-Video and the FastQA implementation were often among the top-three implementations. Tensorflow's MS-SSIM implementation also performs well, lending strong empirical support to the use of MS-SSIM as a training objective for deep networks implemented in Tensorflow.

Since compression forms an important class of distortions encountered in practice, we also report the off-the-shelf performance of SSIM and MS-SSIM implementations on compression distorted data in Table~\ref{tab:comp_default_params}. When restricting the comparisons to compression, the LIBVMAF, Scikit-Image and FastQA implementations still generally outperform other SSIM implementations, while HDRTools and ClearView generally outperform other MS-SSIM implementations.

\subsection{Performance-Efficiency Tradeoffs}
\label{sec:performance_efficiency}
\begin{figure*}[ht]
    \hspace*{\fill}%
    \subfloat[LIVE IQA Database \label{fig:live_iqa_srocc_v_time}]{%
      \includegraphics[width=0.42\linewidth]{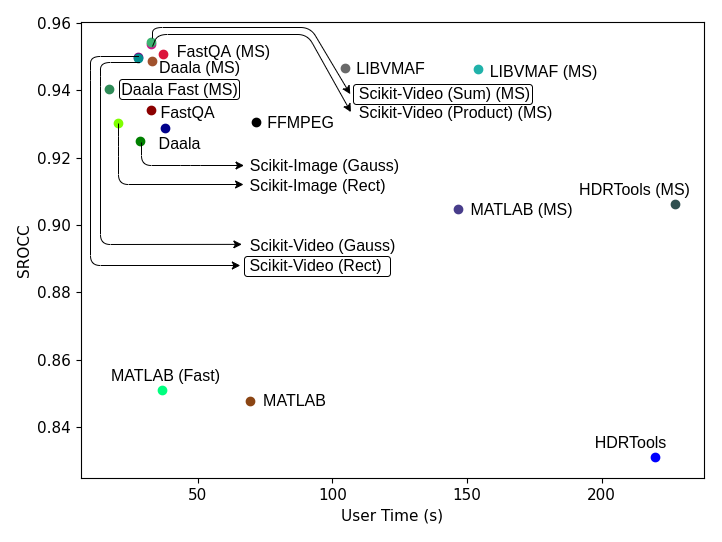}}
    \hfill
    \subfloat[TID 2013 Database \label{fig:tid13_srocc_v_time}]{%
      \includegraphics[width=0.42\linewidth]{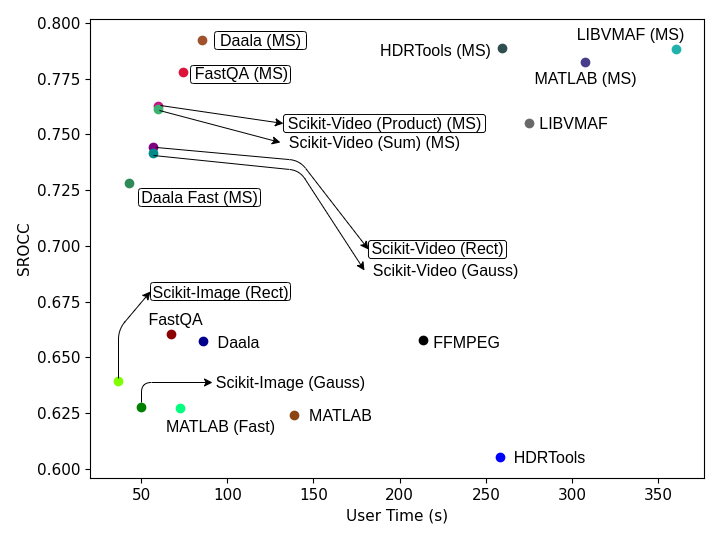}}
    \hspace*{\fill}%
    \\
    \hspace*{\fill}%
    \subfloat[LIVE VQA Database \label{fig:live_vqa_srocc_v_time}]{%
      \includegraphics[width=0.42\linewidth]{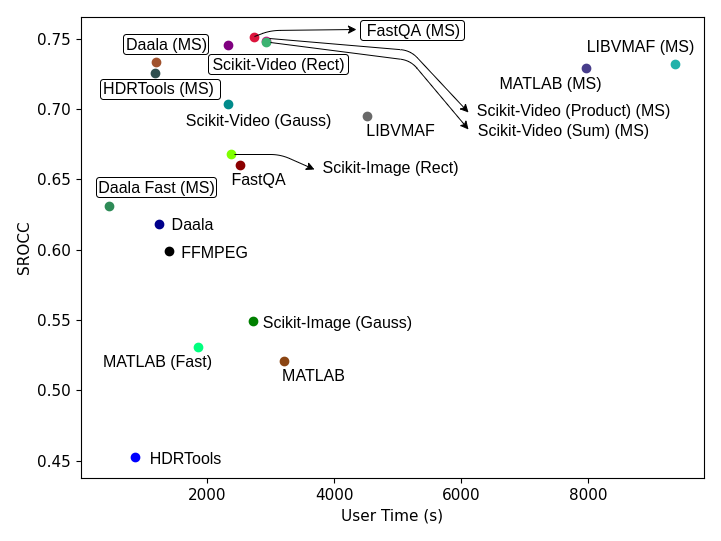}}
    \hfill
    \subfloat[Netflix Public Database \label{fig:nflx_repo_srocc_v_time}]{%
      \includegraphics[width=0.42\linewidth]{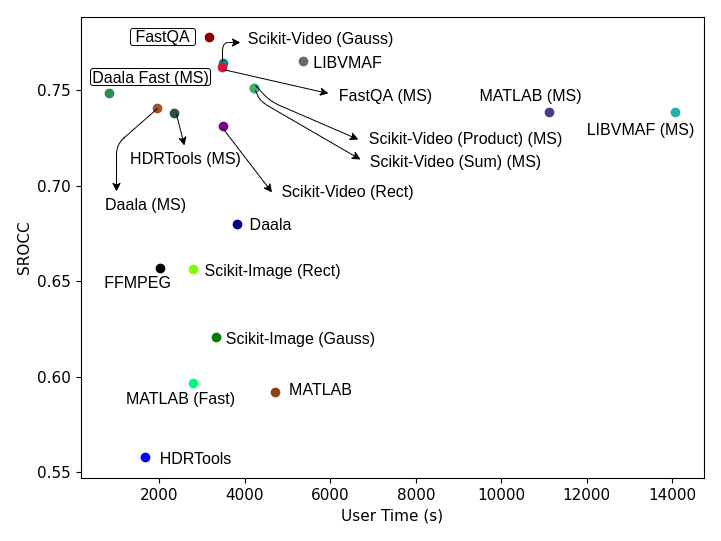}}
    \hspace*{\fill}%
    \\
    \captionsetup{justification=centering}
    \caption{Correlation vs execution time}
    \label{fig:srocc_v_time}
\end{figure*}

In addition to performance (i.e., correlation with subjective scores), it is important to consider the compute efficiency of these implementations. Algorithms that employ sophisticated techniques for down-sampling, calculation of local statistics and multi-scale processing may provide improvements in performance, but often incur the cost of additional computational complexity. When deployed at scale, these additional costs can be significant. 

To evaluate the compared algorithms in the context of this performance-efficiency tradeoff, we plotted the SROCC achieved by each algorithm against their execution time. Since some methods leverage multithreading/multiprocessing, we report the user time instead of the wall time of the processes. 

As with any run-time experiments, we expected to observe slight variations in execution times between runs due to varying system conditions. To account for this, we ran every SSIM implementation on each database five times and recorded the total execution time of each run. We then reported the median run-time over the five runs.

We omitted Tensorflow implementations from these experiments because they run prohibitively slowly on CPUs, and we cannot compare their GPU run-times while all other implementations are run on the CPU. we also omit ClearView implementations because we had to run them on custom hardware. The results on each database are shown in Fig.~\ref{fig:srocc_v_time}, where the Pareto-optimal implementations have been circled.

From these plots, it may be observed that among the implementations that we tested, Daala's Fast MS-SSIM and Scikit-Video's SSIM (using Rectangular windows) implementation are Pareto-optimal most often, followed the FastQA MS-SSIM implementation. In addition, among the SSIM implementations, Daala, LIBVMAF and the FastQA implementations were often Pareto-optimal across databases. Note that while the concept of Pareto-optimality is often used in the context of ``optimizing" an encode in a rate-distortion sense by varying a parameter, no parameters were optimized during our experiments. In our setting, an implementation is considered to be Pareto-optimal among the set of considered implementations if there is no implementation that both achieves a higher SROCC and runs in lesser time.

The nominal computational complexity of SSIM is \(O(MNk^2)\). We propose a method to improve the efficiency of SSIM if the weighting function is rectangular, i.e., \(w(i,j) = 1/k^2\), by using integral images, also known as summed-area tables \cite{ref:int_img_texture} \cite{ref:viola_jones}.

This can be done by forming five integral images as follows:
\begin{equation}
    I^{(1)}_1(i,j) = \begin{cases}
                        \sum\limits_{m\leq i}\sum\limits_{n\leq j} I_1(m,n) & i,j > 0 \\
                        0 & \text{otherwise}
                    \end{cases},
    \label{eq:integral_image_start}
\end{equation}
\begin{equation}
    I^{(1)}_2(i,j) = \begin{cases}
                        \sum\limits_{m\leq i}\sum\limits_{n\leq j} I_2(m,n) & i,j > 0 \\
                        0 & \text{otherwise}
                    \end{cases},
\end{equation}
\begin{equation}
    I^{(2)}_1(i,j) = \begin{cases}
                        \sum\limits_{m\leq i}\sum\limits_{n\leq j} I_1^2(m,n) & i,j > 0 \\
                        0 & \text{otherwise}
                    \end{cases},
\end{equation}
\begin{equation}
    I^{(2)}_2(i,j) = \begin{cases}
                        \sum\limits_{m\leq i}\sum\limits_{n\leq j} I_2^2(m,n) & i,j > 0 \\
                        0 & \text{otherwise}
                    \end{cases},
\end{equation}
\begin{equation}
    I_{12}(i,j) = \begin{cases}
                        \sum\limits_{m\leq i}\sum\limits_{n\leq j} I_1(m,n)I_2(m,n) & i,j > 0 \\
                        0 & \text{otherwise}
                    \end{cases},
\end{equation}

Given the integral image \(I^{(1)}_1\), calculate the sum in any \(k \times k\) window \(\mathcal{W}_{ij}\) in constant time via
\begin{align}
    S^{(1)}_1(i,j) &= I^{(1)}_1(i+k-1,j+k-1) + I^{(1)}_1(i-1,j-1) \nonumber \\
    &- I^{(1)}_1(i+k-1,j-1) - I^{(1)}_1(i-1,j+k-1).
    \label{eq:integral_image_end}
\end{align}
This operation would require \(O(k^2)\) time without the use of an integral image. Similarly, calculate local sums using the other integral images, and denote them \(S^{(1)}_2(i,j)\), \(S^{(2)}_1(i,j)\), \(S^{(2)}_2(i,j)\), and \(S_{12}(i,j)\), respectively. Then calculate the necessary local statistics as
\begin{equation}
    \mu_1(i,j) = S^{(1)}_1(i,j)/k^2,
    \label{eq:mean_from_sum}
\end{equation}
\begin{equation}
    \mu_2(i,j) = S^{(1)}_2(i,j)/k^2,
\end{equation}
\begin{equation}
    \sigma^2_1(i,j) = S^{(2)}_1(i,j)/k^2 - \mu^2_1(i,j),
\end{equation}
\begin{equation}
    \sigma^2_2(i,j) = S^{(2)}_2(i,j)/k^2 - \mu^2_2(i,j),
\end{equation}
\begin{equation}
    \sigma_{12}(i,j) = S_{12}(i,j)/k^2 - \mu_1(i,j)\mu_2(i,j).
    \label{eq:corr_from_sum}
\end{equation}

In this new way of computing rapid SSIM index, which is applicable when using a rectangular SSIM window, the compute complexity of SSIM is reduced to \(O(MN)\).

\section{Scaled SSIM}
\label{sec:scaled_ssim}

\begin{figure}[b]
    \centerline{\includegraphics[width=0.95\linewidth]{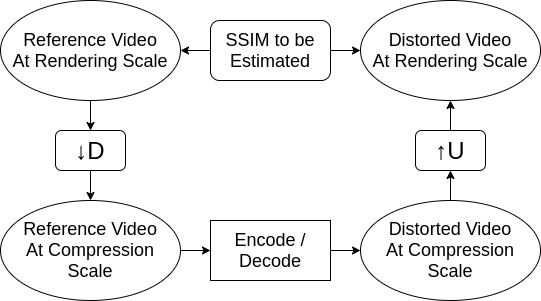}}
    \caption{Video compression pipeline}
    \label{fig:problem_setup}
\end{figure}

Arguably the most widespread use of SSIM (and other picture/video quality models) is in evaluating the quality of compression encodes. On streaming and social media platforms, pictures and videos are commonly encoded at lower resolutions for transmission. This is done either because the source has low-complexity content and can be down-sampled with relatively little additional loss (or if the available bandwidth requires it) or to decrease the decoding load at the user's end. Perceptual distortion models have become common tools for determining the quality of encodes for Rate-Distortion Optimization (RDO) \cite{ref:dyn_optimizer}. Advances in video hardware have enabled the accelerated encoding and decoding of videos, making the distortion estimation step the bottleneck when optimizing encoding ``recipes." From Section~\ref{sec:performance_efficiency}, we know that the computational complexity of SSIM in terms of image dimensions is \(O(MN)\). Including a scale factor \(\alpha\) by which we resize the image, the computational complexity is \(O(\alpha^2MN)\). Due to this quadratic growth, the computational load of distortion estimation is an increasingly relevant issue given the prevalence of high-resolution videos. 

Therefore, it is of great interest to be able to accurately predict the quality of high-resolution videos that are distorted in two steps - scaling followed by compression. For example, consider High Definition (HD) videos that are first resized to a lower resolution, which we call the compression resolution, then encoded and decoded using, for example, H.264 at this compression resolution. The videos are then up-sampled to the original resolution before they are rendered for display. We will refer to this higher resolution as the rendering resolution.

We aim to reduce the computational burden of perceptually-driven RDO by circumventing the computation of SSIM at the rendering resolution, i.e. between the HD source and rendered videos. We propose a suite of algorithms, called Scaled SSIM in \cite{ref:scaled_ssim}, which predict SSIM by only using SSIM values computed at the lower compression resolution during runtime. The video compression pipeline in which we solve the Scaled SSIM problem is illustrated in Fig.~\ref{fig:problem_setup}.

We achieve this using two classes of models that efficiently predict Scaled SSIM, which we refer to as
\begin{itemize}
    \item Histogram Matching
    \item Feature-based models
\end{itemize}
All of the proposed models operate on a per-frame basis.

We first trained and tested the performance of these models on an in-house video corpus of 60 pristine videos. We compressed these videos at 6 compression scales - 144p, 240p, 360p, 480p, 540p, 720p using FFMPEG's H.264 (libx264) encoder at 11 choices of the Quantization Parameter (QP) - \(1, 6, 11, \dots, 51\). In this manner, we obtained a total of 3960 videos having almost 1.75M frames.

On this corpus, we can evaluate the accuracy of predicting SSIM scores, i.e., the correlation between predicted and true SSIM, which was computed using the ``ssim" filter in FFMPEG. However, the end goal is to predict subjective scores, which are not available for this corpus. So, we instead evaluated the performance of our models against subjective scores on the Netflix Public Database.

\subsection{Histogram Matching}

\begin{figure}[t]
    \centerline{\includegraphics[width=\linewidth]{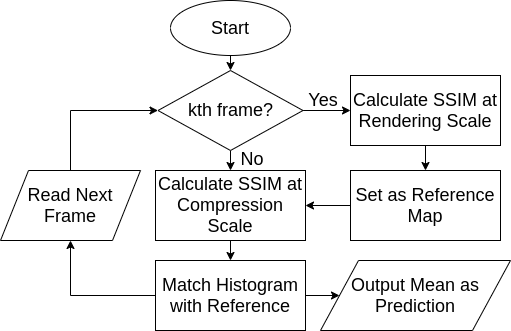}}
    \caption{Histogram Matching Solution}
    \label{fig:hist_match_flowchart}
\end{figure}

We observe a non-linear relationship between SSIM values across encoding resolutions. Because framewise SSIM scores are calculated by averaging the local quality map obtained from SSIM, we can estimate SSIM scores by matching the histograms of these quality maps. However, to match histograms, we require the true histogram at the rendering resolution, which is what we wish to \textit{avoid} estimating.  

So, we instead calculate the ``true" quality map just once every \(k\) frames, and assume that the shape of the true histogram does not change significantly over a short period of \(k-1\) frames. This allows us to reuse this ``reference map" for the next \(k-1\) frames as a heuristic model against which we match the shapes of the next \(k-1\) histograms at the compression scale. This histogram matching algorithm is illustrated in Fig.~\ref{fig:hist_match_flowchart}.

Let \(\alpha \in \left(0,1\right)\) be the factor by which we down-sampled the source video. Then, the ratio of required computation using our proposed approach to SSIM computation directly at the rendered scale is approximately
\begin{equation}
    \left(1 - \frac{1}{k}\right)\alpha^2\left(1 + \beta + \gamma\right) + \frac{1}{k}(1 + \beta).
\end{equation}
    
The factors \(\beta\) and \(\gamma\) account for computing and matching the histograms respectively, which are both \(O(MN)\) operations. This ratio is a decreasing function of \(k\), and approaches \(\alpha^2(1 + \beta + \gamma)\) as \(k \rightarrow \infty\).

By comparison, if the rendered SSIM map were not sampled, the ratio would be approximately \(\alpha^2\). In practice, we have observed that the time taken to compute and match histograms is comparable to the time taken to compute the SSIM map at the compression scale. So, the computational burden of the matching step is small, albeit not negligible.
    
\begin{figure}
    \centerline{\includegraphics[width=\linewidth]{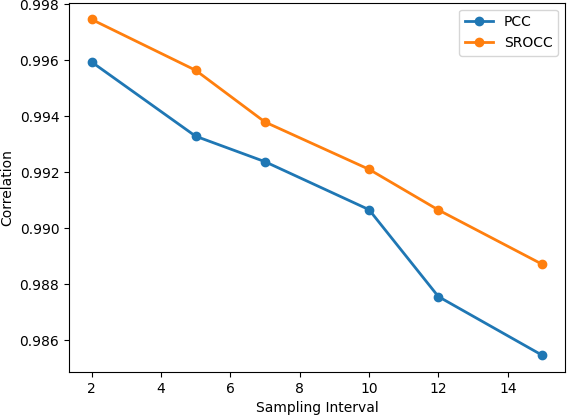}}
    \caption{Correlation vs Sampling interval for Histogram Matching}
    \label{fig:hist_int_results}
\end{figure}

This reduction in computational complexity as \(k\) increases is accompanied by a reduction in performance (accurate prediction of true SSIM), as shown in Fig.~\ref{fig:hist_int_results}. We chose \(k=5\) in all the experiments unless otherwise mentioned.

One drawback of this method is that it requires ``guiding information" in the form of periodically updated reference quality maps. However, this issue is not a factor in the second class of models.

\subsection{Feature-Based Models}
\begin{figure*}[ht]
    \centerline{\includegraphics[width=\textwidth]{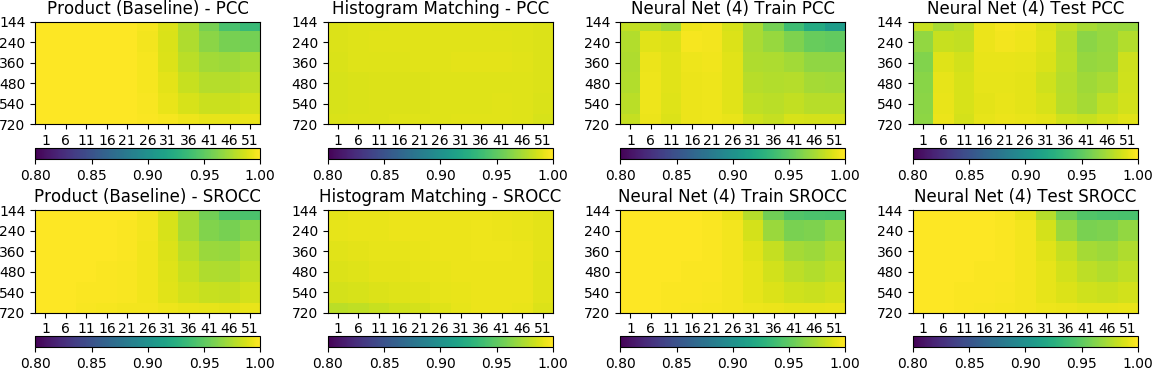}}
    \caption{Variation in performance with choice of Encoding Scale and QP}
    \label{fig:perf_scale_qp}
\end{figure*}

As we observed earlier, the net quality degradation occurs in two steps - scaling and compression. So, we calculate the contribution of each operation and use these as features to calculate the net distortion. 

Let \(X\) be a source video and \(S_\alpha(X)\) denote the video obtained by scaling \(X\) by a factor \(\alpha\). Then, the result of up-sampling the down-sampled video back to the original resolution may be denoted by \(S_\frac{1}{\alpha}(S_\alpha(X))\). The SSIM value between \(X\) and \(S_\frac{1}{\alpha}(S_\alpha(X))\) is a measure of the loss in quality from down-sampling the video. Since this SSIM is independent of the choice of codec and compression parameters, this can be pre-computed.

The second source of quality degradation is compression. Let $C(X;q)$ be the decoded video obtained by encoding the source video $X$ using a Quantization Parameter (QP) $q$. Then, the SSIM value between $S_\alpha(X)$ and $C(S_\alpha(X);q)$ measures the loss of quality resulting from compression of the video.

We use these two SSIM scores as features to predict the true SSIM and refer to these models as Two-Feature Models. In addition, we can also use the scaling factor \(\alpha\) and the QP \(q\) as features. We call such models four-feature models.

In both cases, we train three regressors to predict the SSIM value at the rendering scale on each frame. The three regressors considered are
\begin{itemize}
    \item Linear Support Vector Regressor (Linear SVR)
    \item Gaussian Radial Basis Function SVR (Gaussian SVR)
    \item Fully Connected Neural Network (NN)
\end{itemize}
The Neural Network is a small fully connected network having a single hidden layer with twice the number of neurons as input features. We compared these models to a simple learning-free model, which is used as a baseline. The output of the baseline model is the simple product of the two SSIM features. This is similar to the 2stepQA picture quality model proposed in \cite{ref:2stepqa} \cite{ref:2step_patent}, for two-stage distorted pictures. We call this the Product model.

\subsection{Results}
\begin{table}[t]
    \caption{Correlation with True SSIM on corpus test data}
    \label{tab:corpus_results}
    \begin{center}
        \begin{tabular}{|c|c|c|}
            \hline
            Model & PCC & SROCC \\
            \hline
            NN 2	& 0.9461 & 0.9834 \\
            \textbf{NN 4}	& \textbf{0.9845} & \textbf{0.9869} \\
            Linear SVR 2 & 0.9529 & 0.9759 \\
            Linear SVR 4 & 0.9215 & 0.9201 \\
            Gaussian SVR 2 & 0.8571 & 0.9591 \\
            Gaussian SVR 4	& 0.9598 & 0.9628 \\
            Product (Baseline) & 0.9662	& 0.9829 \\
            \hline
            \textbf{Histogram Matching} & \textbf{0.9933} & \textbf{0.9956} \\
            \hline
        \end{tabular}
    \end{center}
\end{table}
\begin{table}[b]    
    \caption{Correlation with DMOS on Netflix Public Database}
    \label{tab:nflx_repo_results}
    \begin{center}
        \begin{tabular}{|c|c|c|}
            \hline
            Model & PCC & SROCC \\
            \hline
            True SSIM & 0.6962 & 0.6567 \\
            \hline
            \textbf{NN 2} & \textbf{0.6759} & \textbf{0.6425} \\
            Linear SVR 2 & 0.6746 & 0.6196 \\
            Gaussian SVR 2 & 0.6756 & 0.6373 \\
            Product	(Baseline) & 0.6715 & 0.6215 \\
            \hline
            \textbf{Histogram Matching} & \textbf{0.6848} & \textbf{0.6616} \\
            \hline
        \end{tabular}
    \end{center}
    
\end{table}
The correlations between predicted SSIM and true SSIM achieved by the various models on our in-house corpus is shown in Table~\ref{tab:corpus_results}, where ``2" and ``4" denote the number of features input to each learning-based model. Among the feature-based models, four-feature NN performed best. This is to be expected, given the great learning capacity of NNs. 

It is interesting to note, however, that the learning-free product model yielded comparable or better performance at a negligible computational cost. Finally, the Histogram Matching model provided near-perfect predictions, outperforming all other models. The cost of this performance is the additional periodic calculation of reference quality maps/histograms.

Because our corpus contains videos generated at various compression scales and QPs, we were able to evaluate the sensitivity of our best models' performance to these choices of encoding parameters. We illustrate this in Fig.~\ref{fig:perf_scale_qp}. 

We observe that histogram matching performed consistently well across all encoding parameters with only a slight decrease in parameters at high-quality regions, i.e., high compression scale and low QP. We attribute this to the fact that most quality values at low QPs are very close to 1. As a result, the histogram of quality scores at the compression scale is concentrated close to 1, making histogram matching difficult. 

On the other hand, the feature-based models performed poorly in low-quality regions, i.e., low compression scale and high QP. However, videos are seldom compressed at such low qualities, so this does not affect performance in most practical use cases.

Table~\ref{tab:nflx_repo_results} compares models based on the correlation they achieved against subjective opinion scores on the Netflix Public Database. Because our goal is to predict SSIM efficiently, we hold the performance of ``true" SSIM as the gold standard against which we evaluated the performance of the Scaled SSIM models. Because videos in this database were generated by setting bitrates instead of QPs, we only tested our two-feature and histogram matching models. It is important to note that these models were \textbf{not} retrained on the Netflix database. 

From the table, it may be observed that SSIM estimated by Histogram Matching matches the performance of true SSIM. We also observe that the feature-based models approach true SSIM's performance, with the Product Model offering an effective low complexity alternative to the learning-based models.

\section{Device Dependence}
\label{sec:device_dependency}

The Quality of Experience of an individual user varies to a great extent, depending on not only the visual quality of the content, but also various other factors. These include, but may not be limited to \cite{schatz2012impact}:

1. Context of media contents.

2. Users' viewing preferences.

3. Condition of terminal and application used for displaying contents.

4. Network bandwidth and latency.

5. The environment where users view content. (Background lighting conditions, viewing distances, audio devices, etc.)

From the perspective of compression and quality assessment algorithms, the factors to be considered are network fluctuations and terminal screens. The rest of this section restricts the scope to only the size of screens and pixel resolutions, excluding the influence of other variables.

\subsection{Impact of Screen Size}

The issue of screen size on viewing quality originally arose in the context of television \cite{grabe1999role}. Large wall-sized,  theatre(s) and IMAX displays provide better experiences, with subjects reporting increased feelings of 'reality,' i.e. the illusion that viewers are present at the scene. Studies have also shown a possible relation between screen size and viewer's intensity of response to contents. 

\subsection{Displayed Content and Perceived Projection}

The authors of \cite{schatz2012impact} compared the experiences of users of devices of different screen sizes, both on web browsing and video viewing. Mean Opinion Scores were found to vary significantly, with an approximate difference of $\Delta = 1 MOS$ between high-end devices and low-end devices of that time. It was also discovered that viewing of videos on tablets (iPad, etc.) benefited more from displaying contents of higher resolutions (more significant impact on MOS) as compared to mobile phones. User experiences are a combined function of screen size, content resolution, and viewing distance. This is quantified by the contrast sensitivity (CSF) of the HVS \cite{campbell1968application} which is broadly band pass so that increases in resolutions beyond a certain limit are not perceivable, hence pose little impact on user experience. The pass band of the human spatial CSF peaks between 1-10 cycles/degree (cpd) (depending on illumination and temporal factors), falling off rapidly beyond. Naturally, it is desirable that picture and video quality assessment algorithms be able to adapt to screen size against assumed viewing distance, i.e., by characterizing the projection of the screen on the retina \cite{winkler1999issues}. Given a screen height $H$ and a viewing distance $D$, full-screen viewing angle is 
\begin{equation}
    \alpha=2\arctan\left(\frac{H}{2D}\right)
\end{equation}
then, if the viewing screen contains $L$ pixels on each (vertical or horizontal) line, the spatial frequency of the pixel spacing is defined as
\begin{equation}
    f_{max}=\frac{L}{2\alpha}
\end{equation}
in cpd.

\subsection{Transforming across Various Scales}

While primitive specifications of viewing distances and pixel resolutions have been provided by the ITU \cite{bt2020500}, existing subjective picture and video quality databases are mainly defined by their content and their testing environments. Likewise, nearly all quality assessment models that operate over scales use the down-sampling transform
\begin{equation}
    Z_\alpha=\max\left(1,\left\lceil \frac{H_I}{256} \right\rceil\right)
\end{equation}
which creates discontinuities as the height is varied (e.g., $H_1=510$ and $H_2=520$) and does not account for viewing distance D. However, the Self-Adaptive Scale Transform (SAST) \cite{gu2013self} seeks to remediate these weaknesses. SAST uses both the viewing distance and the user's angle of view (including the distance)
\begin{align}
    Z_s & =\sqrt{\frac{H_I\cdot W_I}{H\cdot I}} \\
    & =\sqrt{\frac{1}{4\tan\left(\frac{\theta_H}{2}\right)\tan\left(\frac{\theta_W}{2}\right)}\cdot\frac{H_I}{D}\cdot\frac{W_I}{D}}
\end{align}
where $D$ denotes the viewing distance, $H_I$ and $W_I$ are the height and width of the display, and the corresponding projection on the retina is $H$ by $W$. Commonly assumed, horizontal and vertical viewing angles are $\theta_H=\ang{40}$ and $\theta_W=\ang{50}$.

Because of the band pass nature of the visual system, a picture or video may be preprocessed using Adaptive High-frequency Clipping (AHC) \cite{gu2013adaptive}. In this method, instead of losing high-frequency information during scaling, high frequencies are selectively assigned smaller weights in a wavelet domain.

A dedicated database called VDID2014 \cite{gu2015quality} was created to record human visual responses under varying viewing distances and pixel resolutions. The authors derived an optimal scale selection (OSS) model based on the SAST and AHC model. Instead of directly comparing reference and distorted contents, all frames of the picture or video are first preprocessed according to an assumed or derived viewing distance and pixel resolution, before applying quality assessment algorithms such as SSIM. The OSS model significantly boosted performance of both legacy and modern IQA/VQA models while not significantly increasing computational complexity.

\begin{figure*}[t]
\hspace*{\fill}%
\subfloat[Variation of SSIM performance with window size \label{subfig:ssim_srocc_v_winsize}]{%
\includegraphics[width=\linewidth]{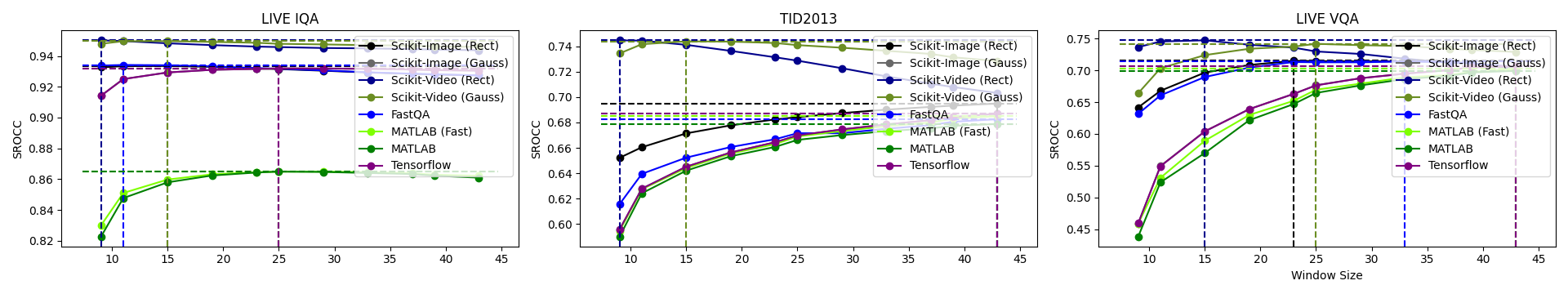}}
\\
\subfloat[Variation of MS-SSIM performance with window size \label{subfig:msssim_srocc_v_winsize}]{%
\includegraphics[width=\linewidth]{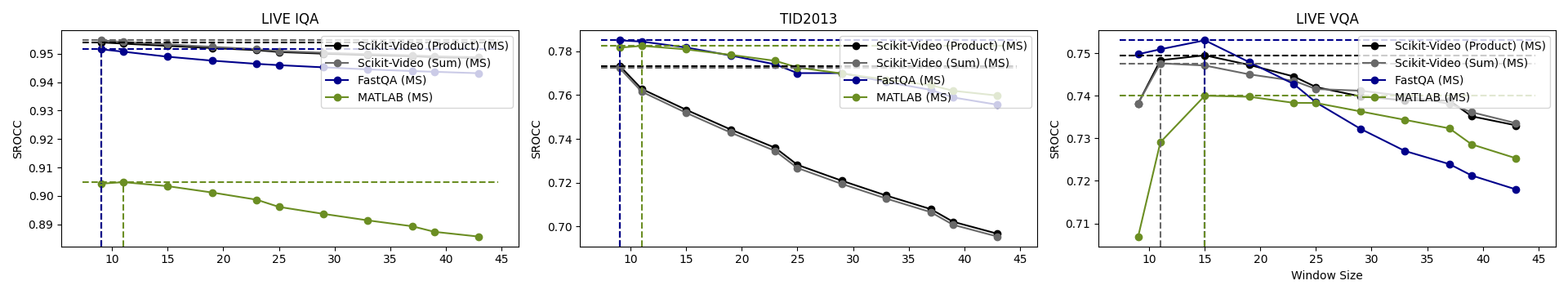}}
\hspace*{\fill}%
\\
\caption{Effect of size and choice of window function on performance}
\label{fig:srocc_v_winsize}
\end{figure*}

\subsection{Tradeoffs}

While modern viewers maintain high expectations of visual quality, regardless of the device and picture or video applications being used, efforts to achieve consistent quality have been inconsistent across screens of various sizes and resolutions. A study of H.264 streams without packet loss \cite{cermak2011relationship} demonstrated that the required bit rate grows linearly with the horizontal screen size, while the expected level of subjective quality was kept fixed.

However, the required bit rates grew much faster with increased expectations of perceived quality, with saturation of MOS at high bit rates, and little quality improvement with increased bit rate. Notwithstanding future improvements in picture/video representation and compression technologies, the results of \cite{cermak2011relationship} supply approximate upper bounds of MOS against content resolution and bit rate.

\subsection{Mobile Devices}

Mobile devices have advanced rapidly in recent years, featuring larger screens and higher resolutions. Most legacy databases that focused on explaining the impact of screen sizes and pixel resolutions were constructed using Personal Digital Assistants (PDA) or older cell phones with displays smaller than 4.5 inches. Popular resolutions of the time of these studies were 320p or 480p, which rarely appear on contemporary mobile devices. In an effort to investigate the same issue on screens larger than 5.7 inches and resolutions of 1440p (2K) or 2160p (4K), a recent subjective study \cite{zou2016perceived} focused on more recent mobile devices having larger resolutions and screen sizes. The contents viewed by the subjects were re-scaled to 4K.

The results from a one-way analysis of variance (ANOVA) suggested no significant relevance between screen size and perceived quality on screens ranging from 4 inches to 5.7 inches. However, a considerable MOS improvement of 0.15 was achieved by 1080p content over 720p content, but no further improvement was observed by increasing the content resolutions to 1440p, suggesting a saturation of perceived quality with resolution. Of course, MOS tends to remain constant across content resolutions higher than that of the display, suggesting that service providers restrict spatial resolutions whenever device specifications are available.

\section{Effect of Window Functions on SSIM}
\label{sec:window}
At the heart of SSIM lies the computation of the local statistics - means, variances, and covariances - comprising the luminance, contrast and structure terms. As described in Section~\ref{sec:background}, the computational complexity of SSIM is \(O(MNk^2)\).

\begin{figure*}[t]
\hspace*{\fill}%
\subfloat[Variation of SSIM performance with window size \label{subfig:comp_ssim_srocc_v_winsize}]{%
\includegraphics[width=0.5\linewidth]{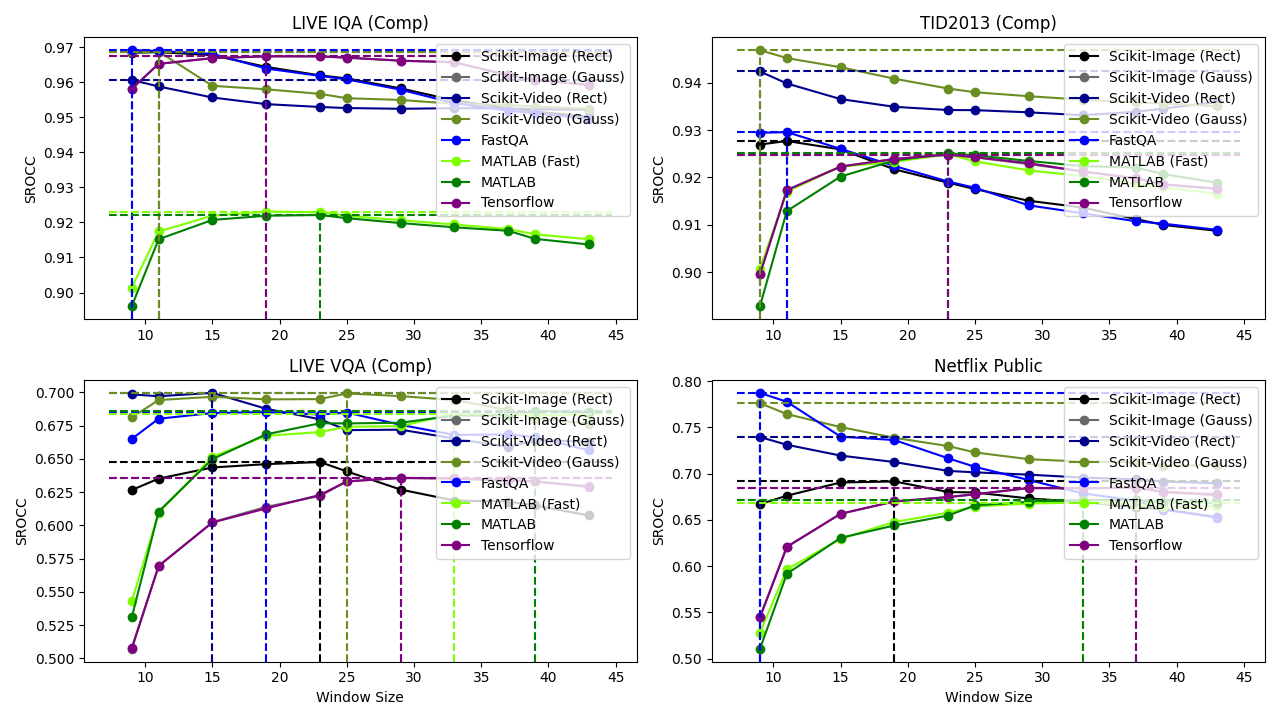}}
\subfloat[Variation of MS-SSIM performance with window size \label{subfig:comp_msssim_srocc_v_winsize}]{%
\includegraphics[width=0.5\linewidth]{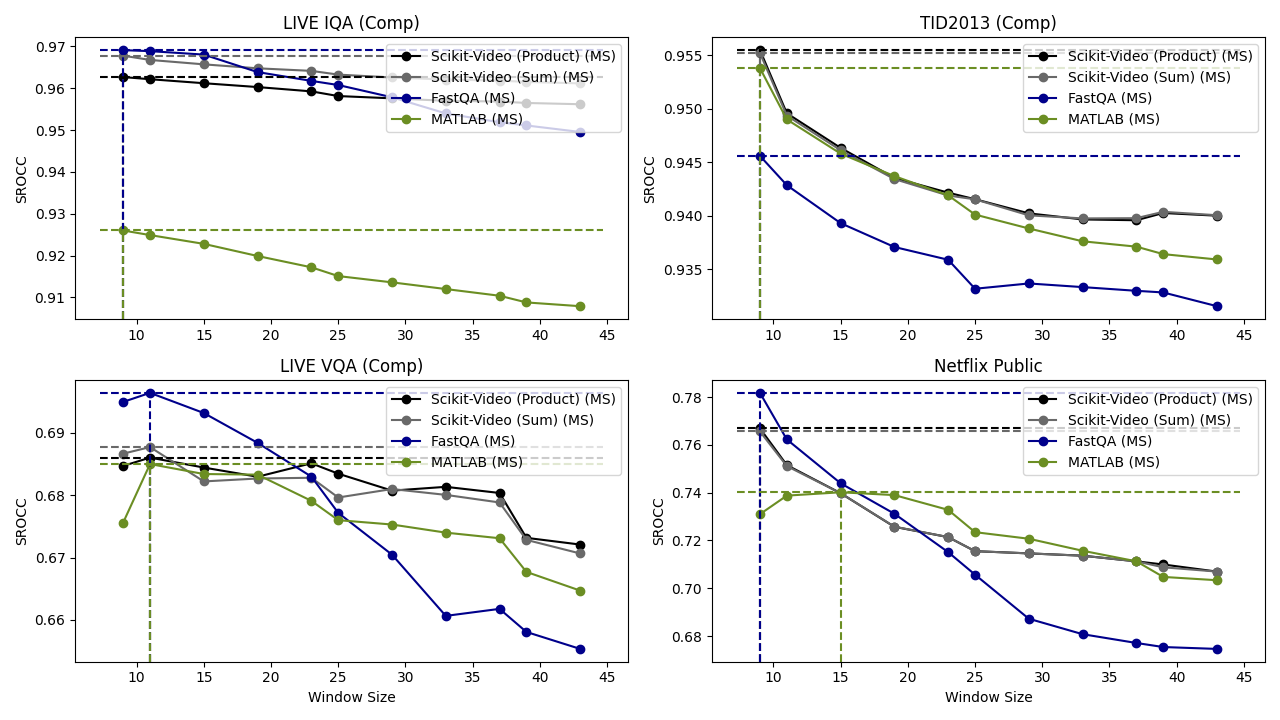}}
\hspace*{\fill}%
\\
\caption{Effect of size and choice of window function on performance on compression data}
\label{fig:comp_srocc_v_winsize}
\end{figure*}

\subsection{Effect of window size}
While computational complexity increases quadratically with window size, using larger windows does not guarantee better performance. Indeed, since picture and video frames are non-stationary, computing local statistics is highly advantageous for local quality prediction. While small values of \(k\) can lead to noisier estimates due to lack of samples, choosing large values of \(k\) risks losing the relevance of local luminance, contrast, structure, and distortion.

As mentioned earlier, the two most common choices of SSIM windows are Gaussian-shaped and rectangular-shaped. Traditionally, the use of rectangular windows is not recommended in image processing, due to frequency side-lobes and resulting ``noise-leakage." Because the frequency response of a rectangular window is a sinc function, undesired high-frequencies can leak into the output. To mitigate this effect, Gaussian filtering is usually preferred, especially for denoising applications or if noise may be present. For these reasons, Gaussian-shaped windows were recommended by the authors of SSIM when calculating local statistics.

To investigate the effect of the choice of window and window size, we considered a set of scaling constant values \(\sigma = 1.0, 1.5, \dots 6.0\) of the Gaussian window, while truncating them at a width of about \(7\sigma\) and forcing the width to be odd. Since only Python implementations allowed setting \(\sigma\), we restricted our experiments to these implementations. All these experiments were conducted using a stride of 1.

Given a Gaussian window of standard deviation \(\sigma\), one can construct analogous rectangular windows in three ways - having the same physical size (i.e., width and height), having the same variance (considering the rectangular window as a sampled uniform distribution), or having the same (3dB) bandwidth. To specify a rectangular window of size \(2K + 1\), we only need to specify \(K\). Equating the variance of the Gaussian to that of a uniform distribution yields \(K = \left\lceil \sigma\sqrt{3}\right\rceil\), where \(\lceil \cdot \rceil\) denotes the ceiling operation. Similarly, equating the 3dB bandwidths of the two filters requires \(K = \left\lceil 1.602\sigma\right\rceil\).

The variation of SSIM performance against window choice is shown in Fig.~\ref{fig:srocc_v_winsize}. For simplicity, we only show the performance of rectangular windows having the same physical size. We observed similar results for rectangular windows having the same variance and 3dB bandwidth. Surprisingly, the figure indicates that rectangular windows are an objectively better design choice than equivalent Gaussian windows! In particular, rectangular windows outperform Gaussian windows for smaller window sizes, achieving slightly higher peak performance. Our experiments suggest that using windows of linear dimension in the range 15 to 20 offers a good tradeoff between performance and computation on both picture databases. Using rectangular windows also offers the possibility of a significant computational advantage, because they can be implemented efficiently using integral images, as discussed in Section~\ref{sec:background}.

We report the variation of performance against window size on compression distorted data in Fig.~\ref{fig:comp_srocc_v_winsize}. From these plots, it may be seen that when tested on compression distortions, SROCC is maximized for smaller window sizes, in the range 7 to 15. In both figures, it may be seen that the Scikit-Video implementations peak at smaller window sizes compared to other implementations. This is explained by the fact that the Scikit-Video implementations downsample images, which increases the ``effective" size of a window function with respect to the original image.

\subsection{Effect of stride}

It is also possible to compute SSIM in a subsampled way, by including a stride \(s\), where \(s\) is the distance between adjacent windows on which SSIM is computed. Then, the computational complexity of SSIM is \(O(MNk^2/s^2)\).

\begin{figure}[ht]
    \centering
    \hspace*{\fill}%
    \subfloat[LIVE IQA Database \label{fig:live_iqa_srocc_v_stride}]{%
       \includegraphics[width=0.48\linewidth]{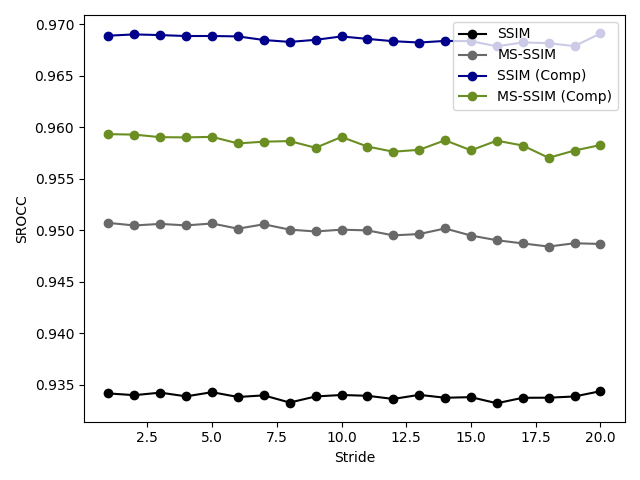}}
    \hfill
    \subfloat[TID 2013 Database \label{fig:tid13_srocc_v_stride}]{%
       \includegraphics[width=0.48\linewidth]{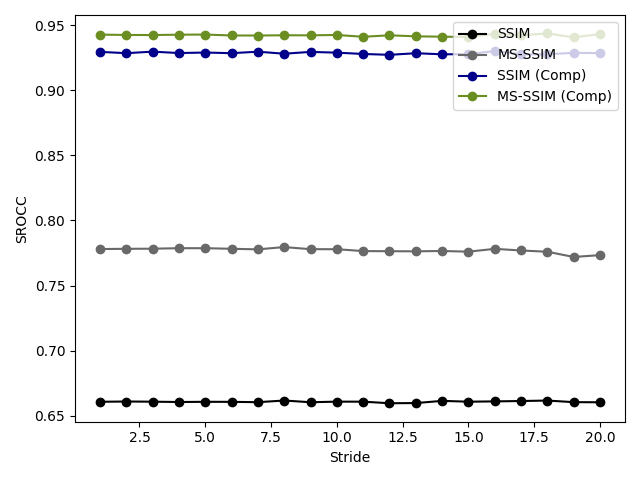}}
    \hspace*{\fill}%
    \\
    \hspace*{\fill}%
    \subfloat[LIVE VQA Database \label{fig:live_vqa_srocc_v_stride}]{%
       \includegraphics[width=0.48\linewidth]{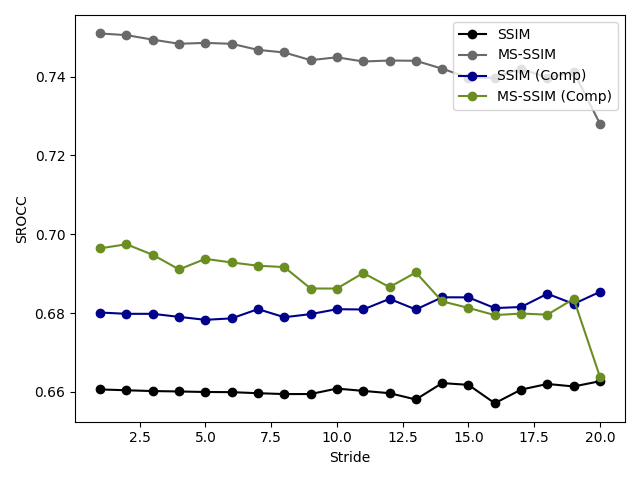}}
    \hfill
    \subfloat[Netflix Public Database \label{fig:nflx_repo_srocc_v_stride}]{%
       \includegraphics[width=0.48\linewidth]{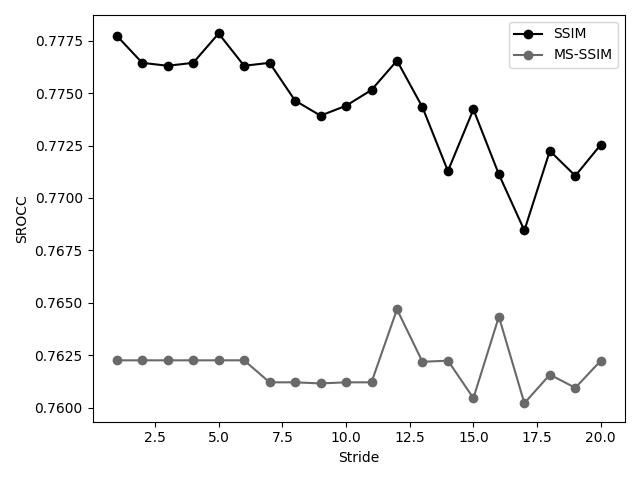}}
    \hspace*{\fill}%
    \\
    \captionsetup{justification=centering}
    \caption{Variation of performance with stride}
    \label{fig:srocc_v_stride}
\end{figure}

We tested the effect of stride on performance using our FastQA python implementation, because none of the existing implementations allow varying the stride. In Fig.~\ref{fig:srocc_v_stride}, we report the variation of SROCC with stride, where lines labelled ``(Comp)" denote the performance on compression distorted data.

From the figure, it may be seen that the SROCC is largely unaffected by stride for \(s \leq 5\). This means that by choosing a stride of  \(s = 5\), we can obtain a significant improvement in efficiency (25x speedup), with little change in prediction performance.

\section{Mapping Scores to Subjective Quality}
\label{sec:nonlinear_mapping}

\begin{table*}[ht]
\caption{Improvement in performance due to linearization}
\label{tab:linear_improv}
\hspace*{\fill}%
\begin{subtable}{0.45\linewidth}
    \centering
    \caption{LIVE IQA Database}
    \begin{tabular}{|c|c|c|c|c|}
    \hline
    \multirow{2}{*}{Metric} & \multicolumn{2}{c|}{PCC} & \multicolumn{2}{c|}{RMSE} \\
    \cline{2-5} 
     & Raw & Fitted & Raw & Fitted \\
    \hline
    ClearView & 0.5928 & 0.7929 & 0.4265 & 0.1161 \\
    \hline
    HDRTools & 0.7002 & 0.8445 & 0.3110 & 0.1021 \\
    \hline
    MATLAB & 0.7311 & 0.8610 & 0.3047 & 0.0969 \\
    \hline
    ClearView (MS) & 0.7677 & 0.8766 & 0.3720 & 0.0917 \\
    \hline
    HDRTools (MS) & 0.6673 & 0.9124 & 0.4576 & 0.0780 \\
    \hline
    \end{tabular}
    \label{tab11}
\end{subtable}
\begin{subtable}{0.45\linewidth}
    \centering
    \caption{TID 2013 Database}
    \begin{tabular}{|c|c|c|c|c|}
    \hline
    \multirow{2}{*}{Metric} & \multicolumn{2}{c|}{PCC} & \multicolumn{2}{c|}{RMSE} \\
    \cline{2-5} 
     & Raw & Fitted & Raw & Fitted \\
    \hline
    ClearView & 0.6612 & 0.7175 & 0.3492 & 0.1239 \\
    % \hline
    % VQMT & 0.652 & 0.692 & 0.272 & 0.128 \\
    \hline
    HDRTools & 0.6205 & 0.6653 & 0.2704 & 0.1327 \\
    \hline
    MATLAB & 0.652 & 0.686 & 0.263 & 0.129 \\
    \hline
    ClearView (MS) & 0.7308 & 0.7508 & 0.3008 & 0.1174 \\
    % \hline
    % VQMT (MS) & 0.768 & 0.819 & 0.357 & 0.102 \\
    \hline
    HDRTools (MS) & 0.7870 & 0.8384 & 0.3738 & 0.0969 \\
    \hline
    \end{tabular}
    \label{tab12}
\end{subtable}
\hspace*{\fill}
\\
\\
\hspace*{\fill}
\begin{subtable}{0.45\linewidth}
    \centering
    \caption{LIVE VQA Database}
    \begin{tabular}{|c|c|c|c|c|}
    \hline
    \multirow{2}{*}{Metric} & \multicolumn{2}{c|}{PCC} & \multicolumn{2}{c|}{RMSE} \\
    \cline{2-5} 
     & Raw & Fitted & Raw & Fitted \\
    \hline
    ClearView & 0.4277 & 0.4625 & 0.4385 & 0.1938 \\
    \hline
    HDRTools & 0.4469 & 0.4789 & 0.3780 & 0.1919 \\
    \hline
    MATLAB & 0.4767 & 0.5595 & 0.3798 & 0.1812 \\
    \hline
    ClearView (MS) & 0.5491 & 0.6569 & 0.4210 & 0.1648 \\
    \hline
    HDRTools (MS) & 0.6701 & 0.7394 & 0.4571 & 0.1472 \\
    \hline
    \end{tabular}
    \label{tab13}
\end{subtable}
\begin{subtable}{0.45\linewidth}
    \centering
    \caption{Netflix Public Database}
    \begin{tabular}{|c|c|c|c|c|}
    \hline
    \multirow{2}{*}{Metric} & \multicolumn{2}{c|}{PCC} & \multicolumn{2}{c|}{RMSE} \\
    \cline{2-5} 
     & Raw & Fitted & Raw & Fitted \\
    \hline
    ClearView & 0.5856 & 0.5896 & 0.4567 & 0.2283 \\
    \hline
    HDRTools & 0.5800 & 0.5800 & 0.3820 & 0.2302 \\
    \hline
    MATLAB & 0.6150 & 0.6150 & 0.3758 & 0.2228 \\
    \hline
    ClearView (MS) & 0.7862 & 0.8119 & 0.4359 & 0.1650 \\
    \hline
    HDRTools (MS) & 0.7171 & 0.7552 & 0.4632 & 0.1852 \\
    \hline
    \end{tabular}
    \label{tab14}
\end{subtable}
\hspace*{\fill}
\end{table*}

\begin{figure}[ht]
    \centering
    \centerline{\includegraphics[width=\linewidth]{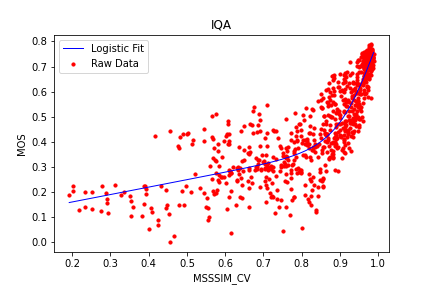}}
    \caption{Example of fitting 5PL curve to a scatter plot of MOS vs. MS-SSIM}
    \label{fig:fig1}
\end{figure}

Because of the different parameter configuration and approximations they use, the many available implementations of SSIM tend to disagree with each other, producing (usually slightly) different scores on the same contents. Of course, any inconsistencies between deployed SSIM models is undesirable, since otherwise in a given application (such as controlling an encoder), changing the SSIM implementation may lead to unpredictable results.

One way to address this issue is by applying a pre-determined function to map the obtained SSIM results to subjective scores. Among a collection of both nonlinear and piecewise linear mappings, the 5PL function in \eqref{eq:five_param_logistic} is particularly useful.

\subsection{Improvement Due to Mapping}

Fig.~\ref{fig:fig1} shows a typical example of fitting raw results (scatter plot of MOS vs. MS-SSIM) to a 5PL function. Both the MOS and objective scores cover fairly wide ranges while the mapping function lies approximately in the middle these.

It can be easily observed that utilizing the fitted curve yields a considerable improvement in PCC and RMSE, while the SROCC remains the same due to the monotonic nature of the function. Table~\ref{tab:linear_improv} shows the improvement obtained in PCC and RMSE after utilizing the fitted curve. For most of the evaluated models there is a considerable performance enhancement.

\subsection{Generalizing to Other Databases}

\begin{table}[b]
\centering
\caption{Generalizability of 5PL SSIM mappings across databases}
\label{tab:5pl_generalization}
\centering
\begin{tabular}{|c|c|c|c|c|}
\hline
\diagbox{SD}{TD} & Netflix Public & LIVE VQA & TID 2013 & LIVE IQA\\
\hline
Netflix Public & 0.228 & 0.197 & 0.307 & 0.718\\
\hline
LIVE VQA & 0.232 & 0.194 & 0.243 & 0.464\\
\hline
TID 2013 & 0.250 & 0.209 & 0.124 & 0.147\\
\hline
LIVE IQA & 0.245 & 0.208 & 0.152 & 0.116\\
\hline
\end{tabular}
\end{table}

While better performance against subjective scores is obtained after mapping the raw data using logistic functions, this only works on databases where subjective scores are available to help optimize the model parameters. In real life, however, social media and streaming service providers lack subjective opinions of their shared of streamed content. This means that it is uncertain whether a set of fitted parameters will apply well to unknown data. In order to study the generalizability provided by logistic mapping, we optimized the function parameters on each individual database, mapped the raw data in the other three databases using the obtained model, as a way of assessing performance on unknown content. If the correlation metrics were to remain similar, it would demonstrate that the logistic function can be used to provided steady performance on unseen data.

The results of the cross-database generalization experiments are shown in Table~\ref{tab:5pl_generalization}. The result in each cell of the table is the RMSE obtained by training a 5PL function on a "source" database (SD), then testing it on each "target" database (TD). From these experiments, we observed that when using MATLAB SSIM, the 5PL function generalized well between the TID 2013 and LIVE IQA Databases.

However, we observed poor generalization when fitting 5PL  on the LIVE VQA database, then testing on the LIVE IQA database. While it may be too much to expect strong performance on video distortions after training on pictures and picture distortions, the lesson learned is still that a user or service provider either select the most relevant database to train on, or conduct a user study directed to their use case, on which a SSIM mapping may be optimized.

\section{Color SSIM}
\label{sec:color_ssim}

Of course, the vast majority of shared and streamed pictures and videos are in color. Hence it is naturally of interest to understand whether SSIM can be optimized to also account for color distortions. however, most available SSIM implementations only operate on the luminance channel. Distortions of the color components may certainly exert considerable influence on subjective quality. The most common approach to incorporating color information into SSIM is to calculate it on each color channel, whether RGB, YUV, or other color space, then combine the channel results.

More sophisticated approaches have been taken to incorporate color channel information into quality models. For example, CMSSIM \cite{hassan2012structural} utilized CIELAB color space \cite{de1978recommendations} distances to better distinguish color distortions and noises. This approach evolved, based on a later subjective study, into CSSIM \cite{hassan2017color}, which generalizes the calculations of SSIM. 

Another approach, called SHSIM \cite{shi2009structure} defines hue similarity (HSIM) much like structural similarity, then combines uses SSIM scores together with the HSIM scores. The combination of two was found to better predict subjective quality than when only using luminance or color.

The method called Quaternion SSIM (QSSIM) \cite{kolaman2011quaternion} combines multi-valued RGB (or any other tristimulus) color space vectors from picture or video pixels into a quaternion representation, providing a formal way to assess luminance and chrominance signals and their degradations together.

Although different in their formulations, these algorithms express individual frames in a tristimulus color space, whether RGB, YUV, or CIELAB depending on the application. In the following, we will define and assess each of these approaches.

\subsection{Quaternion SSIM}
The quaternion SSIM algorithm uses quaternions \cite{kolaman2011quaternion} to represent color pixels with a vector of complex-like components (quaternions are often described as extensions of complex or phasor representations):
\begin{equation}
    q(m,n)=r(m,n)\cdot i+g(m,n)\cdot j+b(m,n)\cdot k.
\end{equation}

The quaternion picture or video frame can then be decomposed into constituent "DC" and "AC" components via
\begin{equation}
    dc\triangleq \mu_{q}=\frac{1}{MN}\sum_{m=1}^{M}\sum_{n=1}^{N} q(m,n),
\end{equation}
and
\begin{equation}
    ac_q\triangleq q(m,n)-\mu_q.
\end{equation}

The quaternion contrast is then defined as
\begin{equation}
    \sigma_q=\sqrt{\frac{1}{(M-1)(N-1)}\sum_{m=1}^{M}\sum_{n=1}^{N}{\left\|ac_q\right\|}_2},
\end{equation}
which, when computed on both reference and test signals, is used to form a correlation factor
\begin{equation}
    \sigma_{q_{ref,dis}}=\frac{1}{(M-1)(N-1)}\sum_{m=1}^{M}\sum_{n=1}^{N}ac_{q_{ref}}\cdot \overline{ac}_{q_{dis}},
\end{equation}
yielding a quaternion formulation similar to the legacy grayscale SSIM:
\begin{equation}
    QSSIM=\left|\left(\frac{2\mu_{q_{ref}}\cdot \mu_{q_{dis}}}{{\mu_{q_{ref}}}^2+{\mu_{q_{dis}}}^2}\right)\left(\frac{\sigma_{q_{ref,dis}}}{{\mu_{q_{ref}}}^2+{\mu_{q_{dis}}}^2}\right)\right|.
\end{equation}

\subsection{CMSSIM}
The CMSSIM algorithm first transforms the input picture or video signal into the CIE XYZ tristimulus color space. These XYZ pixels are then transformed into luminance, red-green, and blue-yellow planes \cite{hassan2017color} as
\begin{equation}
    \left[
    \begin{matrix}
    Q_1 \\ Q_2 \\ Q_3
    \end{matrix}
    \right]
    =
    \left[
    \begin{matrix}
    0.279 & 0.72 & -0.107 \\
    -0.449 & 0.29 & -0.077 \\
    0.086 & -0.59 & 0.501 \\
    \end{matrix}
    \right]
    \left[
    \begin{matrix}
    X \\ Y \\ Z
    \end{matrix}
    \right]
\end{equation}

The resulting chromatic channels are then smoothed using Gaussian kernels, then transformed back into XYZ tristimulus color space via
\begin{equation}
    \left[
    \begin{matrix}
    X \\ Y \\ Z
    \end{matrix}
    \right]
    =
    \left[
    \begin{matrix}
    0.6204 & -1.8704 & -0.1553 \\
    1.3661 & 0.9316 & 0.4339 \\
    1.5013 & 1.4176 & 2.5331 \\
    \end{matrix}
    \right]
    \left[
    \begin{matrix}
    Q_1 \\ Q_2 \\ Q_3
    \end{matrix}
    \right]
\end{equation}
and then into the CIELAB L*, a*, and b*.

The dissimilarities between the reference and test chromatic components is then found as
\begin{align}
    \Delta E(x,y)=&\left(\left(L_1^*(x,y)-L_2^*(x,y)\right)^2 +\nonumber \right. \\
    &\left(a_1^*(x,y)-a_2^*(x,y)\right)^2 +\nonumber \\
    &\left.\left(b_1^*(x,y)-b_2^*(x,y)\right)^2\right)^\frac{1}{2}
\end{align}

which are the used to weight the values of the final SSIM map:
\begin{equation}
    CSSIM=l(x,y) \cdot c(x,y) \cdot s(x,y) \cdot \left(1-\frac{\Delta E(x,y)}{45}\right).
\end{equation}

\subsection{HSSIM}
The HSSIM index is calculated by first transforming pictures or frames into an HSV color space. The color quality is then predicted using a weighted average of SSIM and hue similarity:
\begin{equation}
    HSSIM\left(x,y\right)=\frac{SSIM\left(x,y\right)+0.2 H\left(x,y\right)}{1.2},
\end{equation}
where $H\left(x,y\right)$ is of the same form as SSIM but operates on hue channel instead of grayscale.

Next we discuss straight-forward channel-wise SSIM as applied in YUV space. This model is also tested on the four databases.

\begin{table*}[ht]
\caption{Comparison of color SSIM models}
\label{tab:color_ssim_perf}
\hspace*{\fill}%
\begin{subtable}{0.45\linewidth}
    \centering
    \caption{LIVE IQA Database}
    \label{tab69}
    \begin{tabular}{|c|c|c|c|}
    \hline
    Method & PCC & SRCC & RMSE\\
    \hline
    Baseline SSIM & 0.8594 & 0.8449 & 0.0975 \\
    \hline
    Quaternion (RGB) & 0.8845 & 0.8748 & 0.0889 \\
    \hline
    Quaternion (YUV) & 0.8766 & 0.8657 & 0.0917 \\
    \hline
    Quaternion (LAB) & 0.5983 & 0.5946 & 0.1527 \\
    \hline
    CMSSIM & 0.7873 & 0.7806 & 0.1175 \\
    \hline
    HSSIM & 0.7873 & 0.7809 & 0.1175 \\
    \hline
    FFMPEG & 0.8650 & 0.8507 & 0.0956 \\
    \hline
    Daala (SSIM) & 0.7547 & 0.7163 & 0.1250 \\
    \hline
    Daala (MSSSIM) & 0.8193 & 0.8113 & 0.1093 \\
    \hline
    Daala (FASTSSIM) & 0.7489 & 0.7500 & 0.1263 \\
    \hline
    \textbf{CSSIM} & \textbf{0.8810} & \textbf{0.8790} & \textbf{0.0902} \\
    \hline
    \end{tabular}
\end{subtable}
\begin{subtable}{0.45\linewidth}
    \centering
    \caption{TID 2013 Database}
    \label{tab68}
    \begin{tabular}{|c|c|c|c|}
    \hline
    Method & PCC & SRCC & RMSE\\
    \hline
    Baseline SSIM & 0.6902 & 0.6337 & 0.1287 \\
    \hline
    Quaternion (RGB) & 0.7456 & 0.7155 & 0.1185 \\
    \hline
    \textbf{Quaternion (YUV)} & \textbf{0.7735} & \textbf{0.7564} & \textbf{0.1127} \\
    \hline
    Quaternion (LAB) & 0.3537 & 0.3015 & 0.1663 \\
    \hline
    CMSSIM & 0.6216 & 0.6124 & 0.1393 \\
    \hline
    HSSIM & 0.6217 & 0.6125 & 0.1393 \\
    \hline
    FFMPEG & 0.7168 & 0.6781 & 0.1240 \\
    \hline
    Daala (SSIM) & 0.4691 & 0.4497 & 0.1570 \\
    \hline
    Daala (MSSSIM) & 0.7414 & 0.7472 & 0.1193 \\
    \hline
    Daala (FASTSSIM) & 0.6348 & 0.6037 & 0.1374 \\
    \hline
    CSSIM & 0.6961 & 0.6411 & 0.1277 \\
    \hline
    \end{tabular}
\end{subtable}
\hspace*{\fill}
\\
\\
\hspace*{\fill}
\begin{subtable}{0.45\linewidth}
    \centering
    \caption{LIVE VQA Database}
    \label{tab67}
    \begin{tabular}{|c|c|c|c|}
    \hline
    Method & PCC & SRCC & RMSE\\
    \hline
    Baseline SSIM & 0.5429 & 0.5089 & 0.1836 \\
    \hline
    \textbf{Quaternion (RGB)} & \textbf{0.6709} & \textbf{0.6622} & \textbf{0.1621} \\
    \hline
    Quaternion (YUV) & 0.6139 & 0.5985 & 0.1726 \\
    \hline
    Quaternion (LAB) & 0.2914 & 0.2651 & 0.2091 \\
    \hline
    CMSSIM & 0.4233 & 0.3826 & 0.1980 \\
    \hline
    HSSIM & 0.4194 & 0.3842 & 0.1984 \\
    \hline
    FFMPEG & 0.4651 & 0.4415 & 0.1935 \\
    \hline
    Daala (SSIM) & 0.4702 & 0.4443 & 0.1929 \\
    \hline
    Daala (MS-SSIM) & 0.6488 & 0.6351 & 0.1663 \\
    \hline
    Daala (FastSSIM) & 0.5526 & 0.5208 & 0.1822 \\
    \hline
    CSSIM & 0.6249 & 0.5742 & 0.1707 \\
    \hline
    \end{tabular}
\end{subtable}
\begin{subtable}{0.45\linewidth}
    \centering
    \caption{Netflix Public Database}
    \label{tab66}
    \begin{tabular}{|c|c|c|c|}
    \hline
    Method & PCC & SRCC & RMSE\\
    \hline
    Baseline SSIM & 0.6335 & 0.5904 & 0.2187 \\
    \hline
    Quaternion (RGB) & 0.7621 & 0.7557 & 0.1830 \\
    \hline
    \textbf{Quaternion (YUV)} & \textbf{0.7816} & \textbf{0.7763} & \textbf{0.1763} \\
    \hline
    Quaternion (LAB) & 0.4508 & 0.3690 & 0.2523 \\
    \hline
    CMSSIM & 0.5417 & 0.4379 & 0.2375 \\
    \hline
    HSSIM & 0.5501 & 0.4444 & 0.2360 \\
    \hline
    FFMPEG & 0.6695 & 0.6267 & 0.2099 \\
    \hline
    Daala (SSIM) & 0.6766 & 0.6475 & 0.2081 \\
    \hline
    Daala (MS-SSIM) & 0.7585 & 0.7398 & 0.1842 \\
    \hline
    Daala (FastSSIM) & 0.7695 & 0.7385 & 0.1805 \\
    \hline
    CSSIM & 0.6643 & 0.6056 & 0.2112 \\
    \hline
    \end{tabular}
\end{subtable}
\hspace*{\fill}
\end{table*}

\subsection{Channel-wise SSIM}

While image sensors normally capture RGB data in accordance with photopic (daylight) retinal sampling, most of the structural information is present in the luminance signal. This implies the existence of a lower information (reduced bandwidth) color representation. In fact, both the retinal representation and modern opponent (chromatic) color spaces exploit this property of visual signals. Modern social media and streaming platforms ordinarily process RGB into a chromatic space such as YUV or YCrCb prior to compression. Likewise, IQA/VQA may be defined on color frames represented by luminance and chrominance.

Since chromatic representations reduce the correlation between color planes, the chromatic components with reduced entropies may be down-sampled prior to compression. YCbCr values can be obtained directly from RGB via a linear transformation, typically
\begin{align}
    &Y=0.213\times R+0.715\times G+0.072\times B \\
    &Cb=0.539\times (B-Y) \\
    &Cr=0.635\times(R-Y)
\end{align}

assuming ITU-R BT.709 conversion. However the YCbCr components are defined, a chromatic SSIM model may be defined based on a weighted average of the objective qualities of the individual YCbCr channels
\begin{align}
    F(ref,dis)=&\left(f(Y_{ref},Y_{dis}) + \alpha f(Cb_{ref},Cb_{dis}) + \nonumber \right. \\
    &\left. \beta f(Cr_{ref},Cr_{dis})\right)/(1 + \alpha + \beta)
\end{align}
where \(f(\cdot,\cdot)\) denotes similarity between reference and distorted frames. 

In our case, the baseline SSIM is used as the base QA measure, i.e. \(f(\cdot,\cdot)=SSIM(\text{ref},\text{dis})\). This method is used in popular image and video processing tools like FFMPEG and Daala, where a Color SSIM is calculated as
\begin{equation}
    SSIM=0.8\cdot SSIM^{Y}+0.1\cdot SSIM^{Cb}+0.1\cdot SSIM^{Cr}
\end{equation}

Instead of testing all possible combinations of the hyperparameters during our experiments, we fixed $\alpha=\beta$. We found \(\alpha = \beta = -0.3\) to yield optimal results in our experiments, with the obtained performances of this optimized Color SSIM (CSSIM) on the four databases included in Table~\ref{tab:color_ssim_perf}.

\subsection{Results}

We compared the performances of the above four Color SSIM models on the same databases, with the results tabulated in Table~\ref{tab:color_ssim_perf}. As may be observed, using color information can noticeably boost SSIM's quality prediction power, with QSSIM RGB and YUV yielding the largest gains.

\section{Spatio-temporal Aggregation of Quality Scores}
\label{sec:pool}
In its native form, SSIM is defined on a pair of image regions and returns a local quality score. When applied to a pair of images, a quality map is obtained of (approximately) the same size as the image. The most common method of aggregating these local quality values is to calculate Mean SSIM (MSSIM) to obtain a SSIM score on the entire image. This method of aggregating quality scores is also usually applied when applying SSIM to videos. Frame-wise MSSIM scores are calculated between pairs of corresponding frames, and the average value of MSSIM (over time) is reported as the single SSIM score of the entire video. 

In the context of HTTP streaming, the authors of \cite{ref:http_pool} evaluated various ways of temporal pooling SSIM scores, and found that over longer durations, the simple temporal mean performed about as well as other more sophisticated pooling strategies. Here, we summarize and expand this work by simultaneously testing various spatial and temporal aggregation methods on the two video databases.

We begin by discussing various spatial and temporal pooling strategies that can be used to pool SSIM. Some of these methods require the tuning of hyperparameters. To optimize these hyperparameters, we use the baseline sample mean as the other pooling method, by comparing the SROCC achieved by each choice of hyperparameters. That is, when optimizing a spatial pooling method, we use temporal mean pooling, and vice versa. 

As we will discuss below, many methods have been proposed which leverage either side information, such as visual attention maps, or computationally intensive procedures like optical flow estimation. While these methods offer principled ways to improve the SSIM model, we omit them from our comparisons because they have a high cost, which is often unsuitable for practical deployments of SSIM at large scales.

In subsequent sections, all of the SSIM quality maps were generated using the Scikit-Image implementation of SSIM, with rectangular windows and the default parameters. While the exact results of the experiments may vary slightly with the choice of ``base implementation," we expect these trends to hold across implementations.

\subsection{Moment-Based Pooling}
A straightforward extension of the averaging operation used in SSIM is to replace it by one of the other two Pythagorean means - the Geometric Mean (GM) and the Harmonic Mean (HM). Since local SSIM scores can be negative, we can only use GM and HM pooling on the structural dissimilarity (DSSIM) scores, i.e. \(1 - SSIM\). However, this means that if the SSIM at any location is close to 1, the pooled score is dramatically decreased. So, we do not recommend using GM or HM for spatial pooling. However, framewise SSIM scores are nearly always positive, so we investigate the use of the Pythagorean means for temporal pooling. We also consider the sample median, since it is a robust measure of central tendency (MCT), unlike the mean.
\begin{figure}[t]
    \centering
    \hspace*{\fill}%
    \subfloat[LIVE VQA Database \label{fig:live_vqa_wpool_srocc_v_winsize}]{%
       \includegraphics[width=0.48\linewidth]{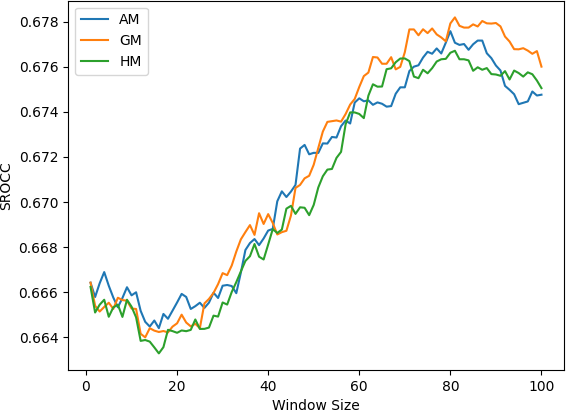}}
    \hfill
    \centering
    \hspace*{\fill}%
    \subfloat[Netflix Public Database \label{fig:nflx_repo_wpool_srocc_v_winsize}]{%
       \includegraphics[width=0.48\linewidth]{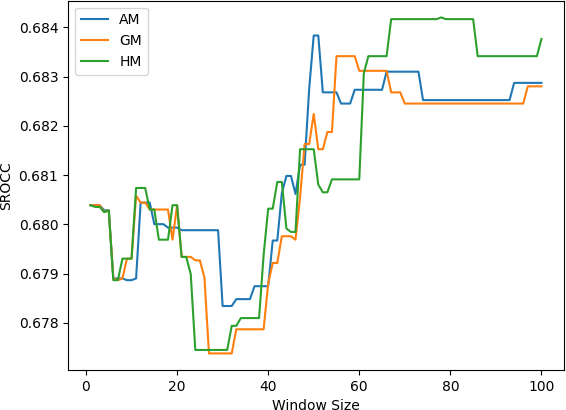}}
    \hfill
    \\
    \captionsetup{justification=centering}
    \caption{Windowed-Moment-Pooling SROCC vs Window Size}
    \label{fig:windowed_pool_srocc_v_winsize}
\end{figure}

Another method of pooling quality scores can be found in the MOVIE index \cite{ref:movie}. It was found that the coefficient of variation (CoV) of quality values correlated well with subjective scores. Let \(\mathbf{x} = (i,j)\) denote spatial indices. Given a quality map \(Q(\mathbf{x}, t)\) having mean value \(\mu_Q(t)\) and standard deviation \(\sigma_Q(t)\), the CoV-pooled score is defined as
\begin{equation}
    S_{\textit{CoV}}(t) = \sigma_Q(t)/\mu_Q(t).
\end{equation}

The same method can be used to pool frame-wise quality scores. One can also adapt the CoV method of temporal pooling using windowing, where the CoV is computed over short temporal windows before being averaged. That is, given a sequence of ``local" temporal CoV values \(\rho_Q(t)\), define the windowed-CoV-pooled scores
\begin{equation}
    S_{\textit{W-CoV}} = \frac{1}{T}\sum_{t} \rho_Q(t).
\end{equation}

In the same vein, windowed versions of the three Pythagorean means can be used for temporal pooling. Using framewise SSIM scores obtained using the Scikit-Image implementation with a rectangular window of size 11, we tested the performance of the three windowed means (W-AM, W-GM, W-HM) and windowed-CoV (W-CoV) pooling. We analyzed the variation of performance of the windowed means against window size, for window sizes \(w = 1, 2, \dots 100\). The results of these experiments are shown in Fig.~\ref{fig:windowed_pool_srocc_v_winsize}. We omitted the W-CoV method from this plot because it gave significantly inferior performance, as shown in Table~\ref{tab:windowed_pool_results}, which lists the best performance of each windowed method.

\begin{table}[t]
    \centering
    \captionsetup{justification=centering}
    \caption{Performance of Windowed-Moment-Pooling}
    \label{tab:windowed_pool_results}
    \begin{tabular}{|c|c|c|c|c|}
        \hline
        Database & Method & PCC & SROCC & RMSE \\
        \hline 
         \multirow{5}{6em}{LIVE VQA} & Baseline SSIM & 0.6645 & 0.6664 & 0.1633 \\
         \cline{2-5}
          & Windowed-AM & 0.6778 & 0.6776 & 0.1607 \\
          \cline{2-5}
          & \textbf{Windowed-GM} & \textbf{0.6788} & \textbf{0.6782} & \textbf{0.1605} \\
          \cline{2-5}
          & \textbf{Windowed-HM} & \textbf{0.6788} & 0.6767 & \textbf{0.1605} \\
          \cline{2-5}
          & Windowed-CoV & 0.4151 & 0.6044 & 0.1988 \\
         \hline
         \multirow{5}{6em}{Netflix Public} & Baseline SSIM & 0.7034 & 0.6804 & 0.2009 \\
         \cline{2-5}
          & Windowed-AM & 0.7222 & 0.6838 & 0.1955 \\
          \cline{2-5}
          & \textbf{Windowed-GM} & \textbf{0.7233} & \textbf{0.6834} & \textbf{0.1951} \\
          \cline{2-5}
          & \textbf{Windowed-HM} & 0.7166 & \textbf{0.6842} & 0.1971 \\
          \cline{2-5}
          & Windowed-CoV & 0.5825 & 0.5997 & 0.2560 \\
         \hline
    \end{tabular}
\end{table}

These plots reveal similar trends. There is an initial decrease in performance with increased window size, but for large enough windows, there is improvement in performance over the baseline. While windowed-CoV performed very poorly, the difference between the three Pythagorean means is small, with windowed-GM being a good choice. However, to observe a reliable improvement in performance, a large window size is needed, of \(k\approx 80\) on the LIVE VQA database and \(k\approx 50\) on the Netflix Public database. However, such large values of \(k\) could lead to significant delays in real-time applications, which may not be a reasonable cost considering the small increase in performance.

\subsection{Five-Number Summary Pooling}
The five-number summary (FNS) \cite{ref:five_number_summary} method was proposed as a better way of summarizing the histogram of a spatial quality map, as compared to the simple mean. Given a spatial quality map \(Q(\mathbf{x},t)\) at time \(t\), let \(Q_{min}(t)\) denote the minimum value, \(Q_1(t)\) denote the 25th percentile, (lower quartile), \(Q_{med}(t)\) denote the median value, \(Q_3(t)\) denote the 75th percentile (upper quartile) and \(Q_{max}(t)\) denote the maximum value. The five number summary is then defined as
\begin{equation}
    S_{\textit{FNS}}(t) = \frac{Q_{min}(t) + Q_1(t) + Q_{med}(t) + Q_3(t) + Q_{max}(t)}{5}
\end{equation}

Of course, FNS may likewise be applied to the framewise quality scores as a way of temporal pooling.

\subsection{Mean-Deviation Pooling}
In \cite{ref:mean_deviation_sim}, the authors proposed a SSIM-like quality index, which is then pooled using a ``mean-deviation" operation. The deviation is defined as the power \(o\) of the Minkowski distance of order \(p\) between the quality values and its mean. More concretely, given a spatial quality map \(Q(\mathbf{x}, t)\) at time \(t\) having mean value \(\mu_Q(t)\), the pooled mean-deviation quality score is given by
\begin{equation}
    S^{(p, o)}_{\textit{MD}}(t) = \left(\left(\frac{1}{MN}\sum_\mathbf{x}(Q(\mathbf{x}, t) - \mu_Q(t))^{p}\right)^{1/p}\right)^o.
\end{equation}

In our experiments, the most common optimal choice was \(p=2\), corresponding to the standard deviation. When applying MD pooling to temporal scores, the final exponent \(o\) does not affect the SROCC, since exponentiation is a monotonic function. So, while we select \(p\) using the SROCC as discussed above, we choose \(o\) for temporal pooling by comparing PCC values.

\subsection{Luminance-Weighted Pooling}
In \cite{ref:ssim_video}, the authors proposed a method of spatial pooling which assigns weights to regions of an image based on the local luminance (brightness), which we call Luminance-Weighted (LW) pooling. These weights are used to account for the fact that the HVS is less sensitive to distortions in dark regions. Following our earlier convention, the local mean \(\mu_1(\mathbf{x})\) is a measure of the local luminance in the reference image. Given a lower limit \(a_s\) and an interval length \(b_s\), the weighting function is defined as
\begin{equation}
    w_{\textit{LW}}(\mathbf{x}) = \begin{cases}
                        0, & \mu_1(\mathbf{x}) < a_s \\
                        (\mu_1(\mathbf{x}) - a_s)/b_s, & a_s \leq \mu_1(\mathbf{x}) < a_s + b_s \\
                        1, & \mu_1(\mathbf{x}) \geq a_s + b_s
                        \end{cases}
\end{equation}

Then, the spatially-weighted SSIM score is given by
\begin{equation}
    S_{\textit{LW}}(t) = \frac{1}{MN}\sum_\mathbf{x} w_{\textit{LW}}(\mathbf{x}) Q(\mathbf{x},t)
\end{equation}

\begin{figure}[t]
    \hspace*{\fill}%
    \subfloat[LIVE IQA Database \label{fig:live_iqa_lw_spat}]{%
       \includegraphics[width=0.48\linewidth]{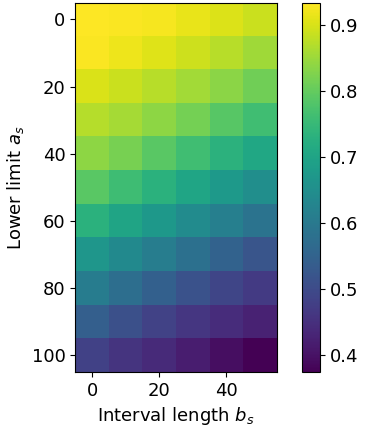}}
    \subfloat[TID2013 Database \label{fig:tid13_lw_spat}]{%
       \includegraphics[width=0.48\linewidth]{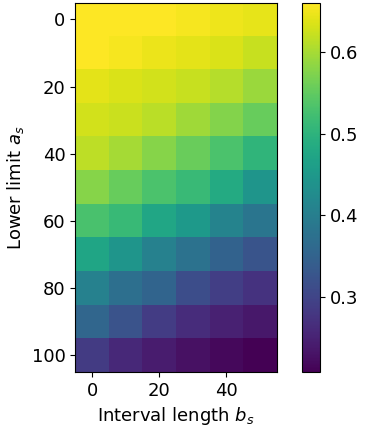}}
    \hspace*{\fill}%
    \\
    \\
    
    \hspace*{\fill}%
    \subfloat[LIVE VQA Database \label{fig:live_vqa_lw_spat}]{%
       \includegraphics[width=0.48\linewidth]{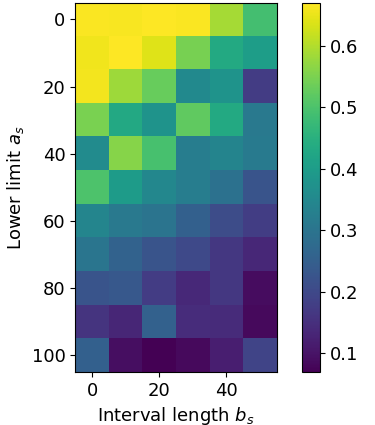}}
    \subfloat[Netflix Public Database \label{fig:nflx_repo_lw_spat}]{%
       \includegraphics[width=0.48\linewidth]{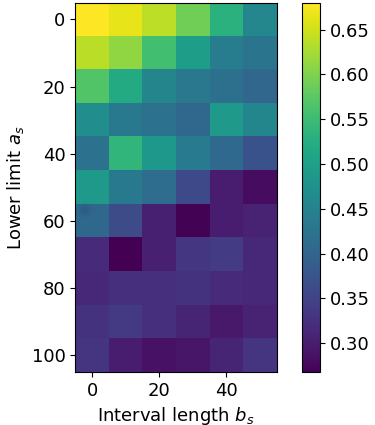}}
    \hspace*{\fill}%
    \captionsetup{justification=centering}
    \caption{LW-Pooling SROCC vs parameters \(a_s\) and \(b_s\)}
    \label{fig:lw_spat}
\end{figure}

We tested the performance of LW-pooling on all four databases for values of \(a_s = 0, 10, \dots, 100\) and \(b_s = 0, 10, \dots 50\). Note that choosing \(a_s=b_s=0\) corresponds to the standard baseline SSIM. The experimental variation of performance (SROCC) against choices of \(a_s\) and \(b_s\) is shown in Fig.~\ref{fig:lw_spat}.

On the LIVE IQA and Netflix Public databases, we observed that the best performance was achieved by the baseline \(a_s=b_s=0\). On the other two databases, the improvement in performance was insignificant, with an elevation of SROCC of less than 0.002. So, we do not recommend using LW-pooling.

\subsection{Distortion-Weighted Pooling}
Distortion-Weighted (DW) pooling is a method that assigns different weights to low and high-quality regions. We consider the common method of distortion weighting, where the weight assigned to a quality score is proportional to a power of the quality score. Concretely, given an exponent \(p\), the spatial DW-pooled score of a spatial quality map \(Q(\mathbf{x}, t)\) as time \(t\) is given by
\begin{equation}
    S_{\textit{DW}}^{(p)}(t) = \frac{\sum\limits_\mathbf{x}\left( 1- Q(\mathbf{x},t)\right)^{p} Q(\mathbf{x}, t)}{\sum\limits_\mathbf{x}\left( 1- Q(\mathbf{x},t)\right)^{p}}.
\end{equation}

Likewise, DW-pooling may be applied to the times series of framewise quality scores to perform temporal DW-pooling. We tested DW-pooling using values of \(p = 1/8, 1/4, \dots 8\) on all four databases, for both spatial and temporal pooling. While DW-pooling can lead to a considerable increase in performance, we also found that the optimal value of \(p\) varied significantly between databases. So, in the absence of a dataset that the user can use to select \(p\) reliably, we do not recommend using DW-pooling off the shelf. If a user does have a relevant dataset, then DW-pooling may be profitably applied. We refer the reader to Table~\ref{tab:ssim_vqa_pool_results} for detailed results.

\subsection{Minkowski Pooling}
The Minkowski Pooling (Mink) method is a generalization of the arithmetic mean, which provides another way to provide additional weight to low quality scores. Because local quality scores can be negative, we pool the DSSIM scores. Given an exponent \(p\), define the spatial Minkowski-pooled score as
\begin{equation}
    S_{\textit{Mink}}^{p}(t) = \frac{1}{MN} \sum\limits_\mathbf{x} \left(1 - Q(\mathbf{x}, t)\right)^p.
\end{equation}

Once again, we tested values of \(p = 1/8, 1/4, \dots, 8\). Note that we omitted \(p=1\) since it is identical to the baseline mean pooling. Spatial Minkowski pooling provided improvement in performance on the video databases, with \(p=4\) being a good choice of \(p\) across databases. However, as with DW-pooling, the optimal choice of \(p\) for temporal pooling varied significantly between databases and any improvement in performance was modest. So, we do not recommend using temporal Minkowski pooling, unless a specific application-relevant dataset is available.

\subsection{Percentile Pooling}

More sophisticated techniques have been proposed to spatially pool of SSIM scores. In \cite{ref:ssim_perc_pool}, the authors propose pooling SSIM scores by visual importance. The visual importance of distortions was measured in two ways: visual attention maps using the Gaze-Attentive Fixation Finding Engine (GAFFE) \cite{ref:gaffe}, and percentile pooling (PP) of quality scores. Because this guide is tailored towards practical application of SSIM, we omit the additional computation of running GAFFE and focus only on PP. 

\begin{figure}[t]
    \hspace*{\fill}%
    \subfloat[LIVE IQA Database \label{fig:live_iqa_pp_spat}]{%
       \includegraphics[width=0.48\linewidth]{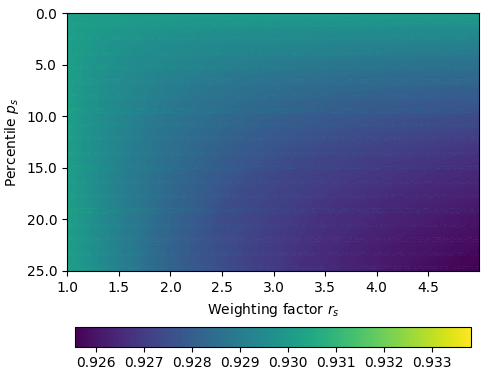}}
    \subfloat[TID2013 Database \label{fig:tid13_pp_spat}]{%
       \includegraphics[width=0.48\linewidth]{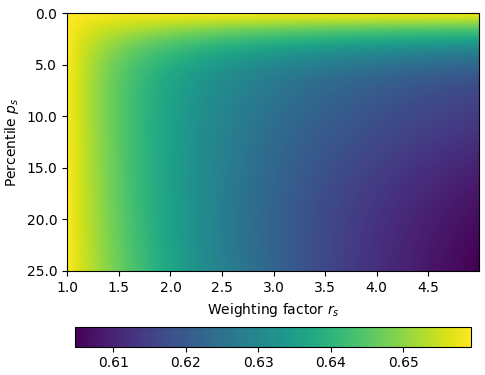}}
    \hspace*{\fill}%
    \\
    \\
    
    \hspace*{\fill}%
    \subfloat[LIVE VQA Database \label{fig:live_vqa_pp_spat}]{%
       \includegraphics[width=0.48\linewidth]{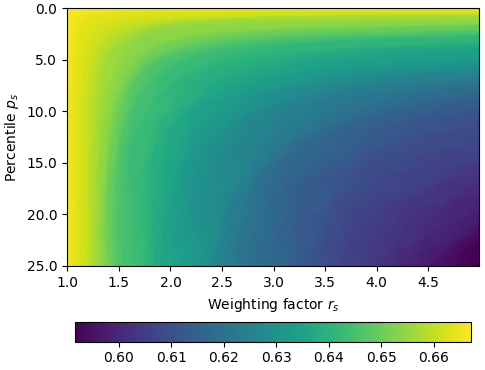}}
    \subfloat[Netflix Public Database \label{fig:nflx_repo_pp_spat}]{%
       \includegraphics[width=0.48\linewidth]{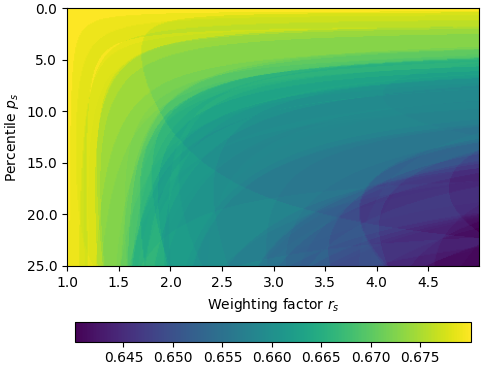}}
    \hspace*{\fill}%
    \captionsetup{justification=centering}
    \caption{Spatial PP SROCC vs parameters \(p_t\) and \(r_t\)}
    \label{fig:pp_spat}
\end{figure}%

The spatial PP method is specified by two parameters: \(p_s\), the percentile of lower values to be modified, and \(r_s\), the factor by which the lower \(p_s\) percentile is weighted. The idea of PP is to heavily weight the worst quality regions, which are likely to heavily bias the perception of quality. Define the lowest \(p_s\) percentile of values of the quality map \(Q(x, t)\) by \(perc(Q, p_s)\). The quality values are then re-weighted as
\begin{equation}
    \tilde{Q}^{(r_s,p_s)}(\mathbf{x}, t) = \begin{cases}
                        Q(\mathbf{x}, t)/r_s, & Q(\mathbf{x}, t) \in perc(Q, p_s) \\
                        Q(\mathbf{x}, t), & otherwise
                      \end{cases}.
\end{equation}

The PP quality score is then defined as the average of the re-weighted quality values:
\begin{equation}
    S^{(r_s,p_s)}_{\textit{PP}}(t) = \frac{1}{MN}\sum_{\mathbf{x}} \tilde{Q}^{(r_s,p_s)}(\mathbf{x}, t).
\end{equation}

Larger values of \(p_s\) penalize more low-quality values, which may dilute the distortion severity, while larger values of \(r_s\) weight the low quality regions more heavily. While the authors of \cite{ref:ssim_perc_pool} were circumspect regarding the value of percentile pooling, they recommended choosing \(p_s=6\) and \(r_s=4000\). However, on all four databases, we found that all choices of parameters \(p_s\) and \(r_s\) performed worse than the baseline. This behavior is illustrated in Fig.~\ref{fig:pp_spat} for values of \(p_s\) in the range \([0, 25]\) (\%) and \(r_s\) in the range \([1,5]\). While the discrepancy between our results and those of \cite{ref:ssim_perc_pool} can be attributed to variations in the implementations, we found that the performance of percentile pooling was inferior across all the tested databases. So, we do not recommend spatial percentile pooling.

\begin{figure}[t]
    \centering
    \hspace*{\fill}%
    \subfloat[LIVE VQA Database \label{fig:live_vqa_pp_temp}]{%
       \includegraphics[width=0.48\linewidth]{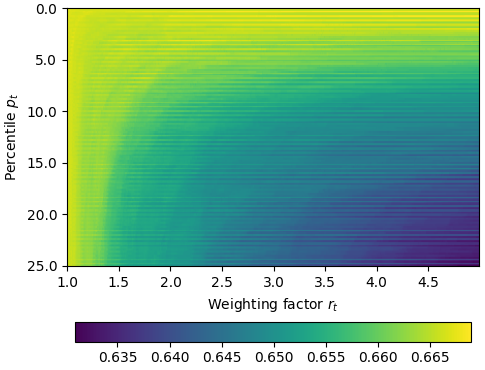}
       \label{fig:pp_temp_live_vqa}}
    \hfill
    \centering
    \hspace*{\fill}%
    \subfloat[Netflix Public Database \label{fig:nflx_repo_pp_temp}]{%
       \includegraphics[width=0.48\linewidth]{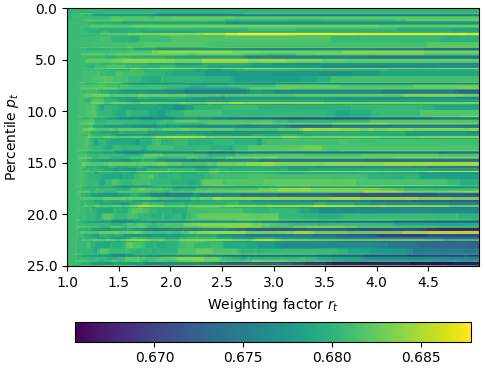}
       \label{fig:pp_temp_nflx_repo}}
       
    \hfill
    \\
    \captionsetup{justification=centering}
    \caption{Temporal PP SROCC vs parameters \(p_t\) and \(r_t\)}
    \label{fig:pp_temp}
\end{figure}%
\begin{table}[t]
    \centering
    \captionsetup{justification=centering}
    \caption{Performance of Temporal PP SSIM}
    \label{tab:pp_temp_results}
    \begin{tabular}{|c|c|c|c|c|}
        \hline
        Database & Method & PCC & SROCC & RMSE \\
        \hline 
         \multirow{2}{6em}{LIVE VQA} & Baseline SSIM & 0.5999 & 0.5971 & 0.1749 \\
         \cline{2-5}
          & \textbf{Temporal PP} & \textbf{0.6163} & \textbf{0.5986} & \textbf{0.1721} \\
         \hline
         \multirow{2}{6em}{Netflix Public} & Baseline SSIM & 0.6815 & 0.6574 & 0.2068 \\
         \cline{2-5}
          & \textbf{Temporal PP} & \textbf{0.6868} & \textbf{0.6601} & \textbf{0.2054} \\
         \hline
    \end{tabular}
\end{table}

We also studied temporal PP of aggregated framewise quality scores by penalizing low-quality frames. Once again, to find the optimal choice of the temporal PP parameters \(p_t\) and \(r_t\), we computed framewise SSIM scores, one which we tested values of \(p_t\) in the range \([0, 25]\) (\%) and \(r_t\) in the range \([1,5]\). Fig.~\ref{fig:pp_temp_live_vqa} and~\ref{fig:pp_temp_nflx_repo} plot the variation of performance (SROCC) with choices of \(p_t\) and \(r_t\) on the LIVE VQA and Netflix Public databases, respectively. The performances of the optimal PP algorithm and baseline SSIM are compared in Table~\ref{tab:pp_temp_results}.

From Table~\ref{tab:pp_temp_results}, it may be observed that temporal PP gave only a minor improvement in performance over baseline SSIM. From the accompanying plots, while there were general trends in performance with variations of each parameter, the prediction performance of the pooled models was sensitive to small perturbations of the parameters. Coupled with the fact that the observed increases in performance were small, it is difficult to reliably identify good choices of the parameters \(p_t\) and \(r_t\). So, we do not recommend using percentile pooling for temporal aggregation either.

\begin{figure}[b]
    \hspace*{\fill}%
    \subfloat[LIVE VQA Database \label{fig:live_vqa_ssim3d}]{%
       \includegraphics[width=0.48\linewidth]{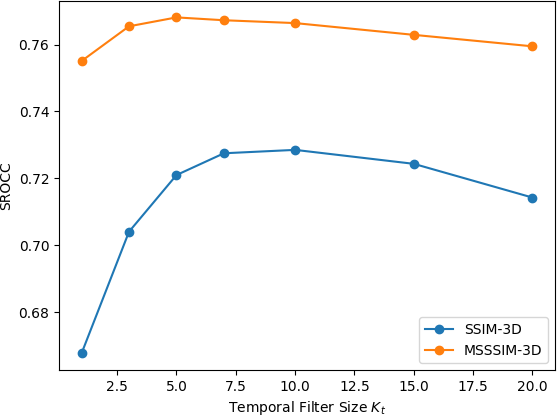}}
    \subfloat[Netflix Public Database \label{fig:nflx_repo_ssim3d}]{%
       \includegraphics[width=0.48\linewidth]{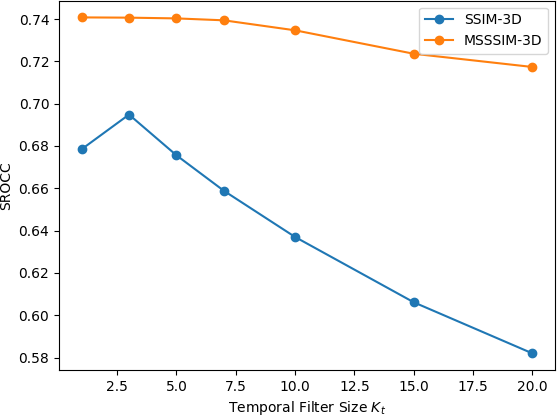}}
    \hspace*{\fill}%
    \captionsetup{justification=centering}
    \caption{Variation of SSIM and MS-SSIM 3D performance with Temporal Window Size \(K_t\)}
    \label{fig:ssim3d_perf}
\end{figure}

\subsection{Spatio-Temporal SSIM}
Efforts have also been made to create spatio-temporal versions of SSIM. In \cite{ref:ssim_3d}, the authors proposed a 3D spatio-temporal SSIM, and its motion-tuned extension. To avoid additional computation related to motion estimation, we consider only the SSIM-3D model and replace the motion-tuned weighting function by a rectangular window. Thus, SSIM-3D was defined as identical to SSIM, other than that local statistics - mean, standard deviation and correlation - were computed on 3D spatio-temporal neighborhoods instead of 2D spatial neighborhoods. Similarly, define MS-SSIM-3D as 3D SSIM computed over multiple spatial scales. Because we choose rectangular windows, we used integral images to efficiently compute the local statistics.

\begin{table}[t]
    \centering
    \captionsetup{justification=centering}
    \caption{Performance of 3D SSIM/MS-SSIM}
    \label{tab:ssim_3d_comp}
    \begin{tabular}{|c|c|c|c|c|}
        \hline
        Database & Method & PCC & SROCC & RMSE \\
        \hline 
         \multirow{4}{6em}{LIVE VQA} & Framewise SSIM & 0.6650 & 0.6677 & 0.1632 \\
         \cline{2-5}
          & \textbf{SSIM 3D} & \textbf{0.7300} & \textbf{0.7285} & \textbf{0.1494} \\
          \cline{2-5}
          & Framewise MS-SSIM & 0.7631 & 0.7551 & 0.1412 \\
         \cline{2-5}
          & \textbf{MS-SSIM 3D} & \textbf{0.7779} & \textbf{0.7681} & \textbf{0.1374} \\
         \hline
         \multirow{4}{6em}{Netflix Public} & Framewise SSIM & 0.7022 & 0.6784 & 0.2012 \\
         \cline{2-5}
          & \textbf{SSIM 3D} & \textbf{0.7086} & \textbf{0.6948} & \textbf{0.1994} \\
          \cline{2-5}
          & Framewise MS-SSIM & 0.7454 & 0.7408 & 0.1884 \\
         \cline{2-5}
          & \textbf{MS-SSIM 3D} & \textbf{0.7512} & \textbf{0.7408} & \textbf{0.1865} \\
         \hline
    \end{tabular}
\end{table}

We tested these 3-D variants of SSIM and MS-SSIM on the LIVE VQA and Netflix Public video databases. We used rectangular filters of size \(11\times11\times K_t\), and investigated the variation of the algorithm's performance with \(K_t\). The performances of the baseline frame-wise SSIM/MS-SSIM (\(K_t = 1\)) models and the best SSIM/MS-SSIM 3D are listed in Table~\ref{tab:ssim_3d_comp}. The variation in performance of SSIM-3D and MS-SSIM-3D with \(K_t\) is plotted in Fig.~\ref{fig:ssim3d_perf}.

From these figures, performance increases by both SSIM-3D and MS-SSIM-3D relative to the 2D frame-based versions may be observed on the LIVE VQA database, with the improvement being much more pronounced in the case of single scale SSIM. When tested on the Netflix-Public database, the improvement was much lower. Again, the improvement was lower for MS-SSIM-3D than SSIM-3D. From the plots, choosing \(K_t\) from the interval \([3, 10]\) offers solid improvement in performance. Another advantage of this approach is that using small rectangular temporal windows, performance increases can be obtained without any increase in computational complexity, by maintaining rolling sums of the last \(K_t\) frames. This can be achieved as below, using a buffer of \(K_t\) frames, leading to an \(O(MNK_t)\) memory complexity.

As in equations \eqref{eq:mean_from_sum} - \eqref{eq:corr_from_sum}, we can calculate the local statistics from the analogously defined sums over 3D neighborhoods \(S_1^{(1)}(i,j,k)\), \(S_1^{(2)}(i,j,k)\), \(S_2^{(1)}(i,j,k)\), \(S_2^{(2)}(i,j,k)\), and \(S_{12}(i,j,k)\). As an illustrative example, consider
\begin{equation}
    S_1^{(1)}(i,j,k) = \sum_{m = i-l+1}^{i} \sum_{n = j-l+1}^{j} \left(\sum_{o = k-K_t+1}^{k} I_1(m,n,o)\right).
    \label{eq:3d_im1_mean_sum}
\end{equation}

Defining the temporal sum
\begin{equation}
    T_1^{(1)}(i,j,k) = \left(\sum_{o = k-K_t+1}^{k} I_1(i,j,o)\right),
\end{equation}
we can rewrite \eqref{eq:3d_im1_mean_sum} as
\begin{equation}
    S_1^{(1)}(i,j,k) = \sum_{m = i-l+1}^{i} \sum_{n = j-l+1}^{j} T_1^{(1)}(m,n,k).
\end{equation}

\begin{table*}[!ht]
% \hspace*{1em}%
\captionsetup{justification=centering}
\caption{Comparing the Performance of Spatial Pooling Methods on Image Databases}
\label{tab:ssim_iqa_pool_results}
\begin{subtable}{0.45\linewidth}
    \centering
    \captionsetup{justification=centering}
    \subcaption{LIVE IQA Database}
    \label{subtab:all_live_iqa_results}
    \begin{tabular}{|c|c|c|c|}
        \hline
         Method & PCC & SROCC & RMSE \\
         \hline
         AM & 0.944	& 0.934	& 0.091\\
         \hline
         CoV & 0.940	& 0.931	& 0.093\\
         \hline
         \(\text{MD}^{(2,1)}\) & 0.906	& 0.887	& 0.116\\
         \hline
         FNS & 0.939	& 0.929	& 0.0941\\
         \hline
         \(\text{DW}^{(1/4)}\)	& 0.944	& 0.934	& 0.090 \\
         \hline
        \(\mathbf{\textbf{Mink}^{(2)}}\)	& \textbf{0.944}	& \textbf{0.934}	& \textbf{0.090} \\
        \hline
    \end{tabular}
\end{subtable}
\hspace*{4em}%
\begin{subtable}{0.45\linewidth}
    \centering
    \captionsetup{justification=centering}
    \caption{TID 2013 Database}
    \label{subtab:all_tid13_results}
    \begin{tabular}{|c|c|c|c|}
        \hline
         Method & PCC & SROCC & RMSE \\
         \hline
         AM & 0.711	& 0.659	& 0.125 \\
         \hline
         CoV & 0.741	& 0.718	& 0.119 \\
         \hline
         \(\text{MD}^{(2,1)}\) & 0.687	& 0.705	& 0.129 \\
         \hline
         FNS & 0.692	& 0.583	& 0.128 \\
         \hline
         \(\text{DW}^{(1/8)}\) & 0.732 &	0.715 & 0.121 \\
         \hline
        \(\mathbf{\textbf{Mink}^{(4)}}\) & \textbf{0.755} & \textbf{0.747} & \textbf{0.117} \\
        \hline
    \end{tabular}
\end{subtable}
\end{table*}

\begin{table*}[!ht]
\captionsetup{justification=centering}
\caption{Comparing the Performance of Pairs of Spatial and Temporal Pooling Methods on Video Databases}
\label{tab:ssim_vqa_pool_results}
\begin{subtable}{0.45\linewidth}
    \centering
    \captionsetup{justification=centering}
    \caption{LIVE VQA Database - PCC}
    \label{subtab:all_live_vqa_pccs}
    \begin{tabular}{|c|c|c|c|c|c|c|}
        \hline
        \backslashbox{TP}{SP} & AM & CoV & \(\text{MD}^{(2,3)}\) & FNS & \(\text{DW}^{(1)}\) & \(\text{Mink}^{(4)}\) \\
        \hline
        AM & 0.664 & 0.766 & 0.708 & 0.580 & 0.770 &	0.781\\
        \hline
        GM & 0.665	& 0.738 & 0.650 & 0.522 & 0.777 & 0.757\\
        \hline
        HM & 0.661	&0.709	&0.435	&0.518 & 0.781	& 0.736\\
        \hline
        CoV & 0.521 & 0.221 & 0.153 &	0.341 & 0.578	& 0.161\\
        \hline
        \(\text{MD}^{(2,3)}\) & 0.587	&	0.668	&	0.695	&	0.530 & 0.622	& 0.580\\
        \hline
        FNS & 0.647	& 0.726	& 0.641	& 0.510 & 0.775	& \textbf{0.784}\\
        \hline
        \(\text{W-AM}^{(80)}\) & 0.677	& 0.770	& 0.699	& 0.540 & 0.773	& \textbf{0.782}\\
        \hline
        \(\text{W-GM}^{(81)}\) & 0.678	& 0.746	& 0.651	& 0.541 & 0.780	& 0.761\\
        \hline
        \(\text{W-HM}^{(81)}\) & 0.643	& 0.725	& 0.615	& 0.541 & \textbf{0.784}	& 0.743 \\
        \hline
        \(\text{DW}^{(2)}\) & 0.694	& 0.762	& 0.709	& 0.611 & 0.767	& 0.753 \\
        \hline
        \(\text{Mink}^{(2)}\) & 0.631	& 0.770	& \textbf{0.788}	& 0.523 & \textbf{0.783}	& 0.780\\
        \hline
    \end{tabular}
\end{subtable}
\hspace*{4em}%
\begin{subtable}{0.45\linewidth}
    \centering
    \captionsetup{justification=centering}
    \caption{Netflix Public Database - PCC}
    \label{subtab:all_nflx_repo_pccs}
    \begin{tabular}{|c|c|c|c|c|c|c|c|c|c|}
        \hline
        \backslashbox{TP}{SP} & AM & CoV & \(\text{MD}^{(4,1)}\) & FNS & \(\text{DW}^{(8)}\) & \(\text{Mink}^{(8)}\)\\
        \hline
        AM & 0.703	& 0.802	& 0.887	& 0.677 & \textbf{0.924}	& 0.885\\
        \hline
        GM & 0.699	& 0.814	& 0.892	& 0.679 & 0.816	& 0.890\\
        \hline
        HM & 0.702	& 0.820	& 0.897	& 0.679 & 0.810	& 0.894\\
        \hline
        CoV & 0.526	& 0.111	& 0.486 &	0.529 & 0.532	& 0.457\\
        \hline
        \(\text{MD}^{(2,1)}\) & 0.570	& 0.495	& 0.277	& 0.511 & 0.292	& 0.323\\
        \hline
        FNS & 0.689	& 0.786	& 0.885	& 0.671 & \textbf{0.917}	& 0.881\\
        \hline
        \(\text{W-AM}^{(50)}\) & 0.718	& 0.799	& 0.880	& 0.682 & \textbf{0.921}	& 0.882\\
        \hline
        \(\text{W-GM}^{(55)}\) & 0.719	& 0.804	& 0.882	& 0.684 & 0.866	& 0.884\\
        \hline
        \(\text{W-HM}^{(78)}\) & 0.704	& 0.809	& 0.884	& 0.684 & 0.842	& 0.889 \\
        \hline
        \(\text{DW}^{(1/4)}\) & 0.698	& 0.804	& 0.888	& 0.680 & \textbf{0.923}	& 0.887 \\
        \hline
        \(\text{Mink}^{(1/4)}\) & 0.710	& 0.801	& 0.886	& 0.668 & \textbf{0.901}	& 0.875\\
        \hline
    \end{tabular}
\end{subtable}
\\
\\
\begin{subtable}{0.45\linewidth}
    \centering
    \captionsetup{justification=centering}
    \caption{LIVE VQA Database - SROCC}
    \label{subtab:all_live_vqa_sroccs}
    \begin{tabular}{|c|c|c|c|c|c|c|c|c|c|}
        \hline
        \backslashbox{TP}{SP} & AM & CoV & \(\text{MD}^{(2,3)}\) & FNS & \(\text{DW}^{(1)}\) & \(\text{Mink}^{(4)}\)\\
        \hline
        AM & 0.667	& 0.762	& \textbf{0.788}	& 0.584 & 0.749	& 0.762\\
        \hline
        GM & 0.666	& 0.727	& 0.726	& 0.579 & 0.758	& 0.730\\
        \hline
        HM & 0.660	& 0.695	& 0.552	& 0.574 & 0.764	& 0.706\\
        \hline
        CoV & 0.599	& 0.039	& 0.119	& 0.532 & 0.656	& 0.153\\
        \hline
        \(\text{MD}^{(2,3)}\) & 0.572	& 0.669	& 0.728	& 0.540 & 0.610	& 0.557\\
        \hline
        FNS & 0.640	& 0.752	& 0.741	& 0.556 & 0.763	& 0.778\\
        \hline
        \(\text{W-AM}^{(80)}\) & 0.678	& 0.764	& \textbf{0.791}	& 0.592 & 0.754	& 0.764\\
        \hline
        \(\text{W-GM}^{(81)}\) & 0.678	& 0.731	& 0.731	& 0.594 & 0.763	& 0.736\\
        \hline
        \(\text{W-HM}^{(81)}\) & 0.677	& 0.699 & 0.672	& 0.593 & 0.766	& 0.708 \\
        \hline
        \(\text{DW}^{(2)}\) & 0.693	& 0.743	& \textbf{0.790}	& 0.564 & 0.766	& 0.726 \\
        \hline
        \(\text{Mink}^{(2)}\) & 0.663	& 0.754	& \textbf{0.789}	& 0.581 & 0.775	& 0.754\\
        \hline
    \end{tabular}
\end{subtable}
\hspace*{4em}%
\begin{subtable}{0.45\linewidth}
    \centering
    \captionsetup{justification=centering}
    \caption{Netflix Public Database - SROCC}
    \label{subtab:all_nflx_repo_sroccs}
    \begin{tabular}{|c|c|c|c|c|c|c|c|}
        \hline
        \backslashbox{TP}{SP} & AM & CoV & \(\text{MD}^{(4,1)}\) & FNS & \(\text{DW}^{(8)}\) & \(\text{Mink}^{(8)}\)\\
        \hline
        AM & 0.680	& 0.768	& 0.871	& 0.633 & \textbf{0.911}	& 0.872\\
        \hline
        GM & 0.682	& 0.770	& 0.872	& 0.634 & 0.837	& 0.874\\
        \hline
        HM & 0.681	& 0.773	& 0.875	& 0.635 & 0.832	& 0.877\\
        \hline
        CoV & 0.479	& 0.143	& 0.437	& 0.497 & 0.610	& 0.409\\
        \hline
        \(\text{MD}^{(2,1)}\) & 0.445	& 0.437	& 0.104	& 0.463 & 0.246	& 0.228\\
        \hline
        FNS & 0.667	& 0.760	& 0.868	& 0.629 & \textbf{0.886}	& 0.866\\
        \hline
        \(\text{W-AM}^{(50)}\) & 0.684	& 0.765	& 0.863	& 0.632 & \textbf{0.896}	& 0.868\\
        \hline
        \(\text{W-GM}^{(55)}\) & 0.683	& 0.764	& 0.865	& 0.633 & 0.863	& 0.869\\
        \hline
        \(\text{W-HM}^{(78)}\) & 0.684	& 0.765	& 0.868	& 0.633 & 0.839	& 0.872\\
        \hline
        \(\text{DW}^{(8)}\) & 0.683	& 0.768	& 0.870	& 0.634	& \textbf{0.910}	& 0.873\\
        \hline
        \(\text{Mink}^{(8)}\) & 0.682	& 0.769	& 0.871	& 0.633	& \textbf{0.913}	& 0.869\\
        \hline
    \end{tabular}
\end{subtable}
\\
\\
\begin{subtable}{0.45\linewidth}
    \centering
    \captionsetup{justification=centering}
    \caption{LIVE VQA Database - RMSE}
    \label{subtab:all_live_vqa_rmses}
    \begin{tabular}{|c|c|c|c|c|c|c|c|}
        \hline
        \backslashbox{TP}{SP} & AM & CoV & \(\text{MD}^{(2,3)}\) & FNS & \(\text{DW}^{(1)}\) & \(\text{Mink}^{(4)}\)\\
        \hline
        AM & 0.163	& 0.140	& 0.154	& 0.178 & 0.140	& 0.137\\
        \hline
        GM & 0.163	& 0.148	& 0.166	& 0.186 & 0.138	& 0.143\\
        \hline
        HM & 0.164	& 0.154	& 0.197	& 0.187 & 0.137	& 0.148\\
        \hline
        CoV & 0.187	& 0.213	& 0.216	& 0.205 & 0.178	& 0.216\\
        \hline
        \(\text{MD}^{(2,3)}\) & 0.177	& 0.163	& 0.157 &	0.185 & 0.171	& 0.178\\
        \hline
        FNS & 0.167	& 0.150	& 0.168	& 0.188 & 0.138	& \textbf{0.136} \\
        \hline
        \(\text{W-AM}^{(80)}\) & 0.161	& 0.139	& 0.156	& 0.184 & 0.139	& \textbf{0.136}\\
        \hline
        \(\text{W-GM}^{(81)}\) & 0.161	& 0.146	& 0.166	& 0.184 & 0.137	& 0.142\\
        \hline
        \(\text{W-HM}^{(81)}\) & 0.167	& 0.151	& 0.172	& 0.184 & \textbf{0.136}	& 0.146\\
        \hline
        \(\text{DW}^{(2)}\) & 0.157	& 0.142	& 0.154	& 0.173	& 0.140	& 0.144\\
        \hline
        \(\text{Mink}^{(2)}\) & 0.170	& 0.139	& \textbf{0.135}	& 0.186	& \textbf{0.136}	& 0.137\\
        \hline
    \end{tabular}
\end{subtable}
\hspace*{4em}%
\begin{subtable}{0.45\linewidth}
    \centering
    \captionsetup{justification=centering}
    \caption{Netflix Public Database - RMSEs}
    \label{subtab:all_nflx_repo_rmses}
    \begin{tabular}{|c|c|c|c|c|c|c|c|}
        \hline
        \backslashbox{TP}{SP} & AM & CoV & \(\text{MD}^{(4,1)}\) & FNS & \(\text{DW}^{(8)}\) & \(\text{Mink}^{(8)}\)\\
        \hline
        AM & 0.201	& 0.169	& 0.130	& 0.208 & \textbf{0.108}	& 0.132\\
        \hline
        GM & 0.202	& 0.164	& 0.128	& 0.208 & 0.163	& 0.129\\
        \hline
        HM & 0.201	& 0.162	& 0.125	& 0.208 & 0.166	& 0.127\\
        \hline
        CoV & 0.240	& 0.281	& 0.247	& 0.240 & 0.239	& 0.251\\
        \hline
        \(\text{MD}^{(2,1)}\) & 0.232	& 0.246	& 0.272	& 0.243 & 0.270	& 0.268\\
        \hline
        FNS & 0.205	& 0.175	& 0.131	& 0.210 & \textbf{0.113}	& 0.134\\
        \hline
        \(\text{W-AM}^{(50)}\) & 0.197	& 0.170	& 0.134	& 0.206 & \textbf{0.110}	& 0.133\\
        \hline
        \(\text{W-GM}^{(55)}\) & 0.196	& 0.168	& 0.133	& 0.206 & 0.141	& 0.132\\
        \hline
        \(\text{W-HM}^{(78)}\) & 0.200	& 0.166	& 0.132	& 0.206 & 0.152	& 0.130\\
        \hline
        \(\text{DW}^{(1/4)}\) & 0.202	& 0.168	& 0.130 & 0.207	& \textbf{0.109}	& 0.130\\
        \hline
        \(\text{Mink}^{(1/4)}\) & 0.199	& 0.169	& 0.131	& 0.210	& \textbf{0.123}	& 0.137\\
        \hline
    \end{tabular}
\end{subtable}
\end{table*}

\begin{table*}[!ht]
\captionsetup{justification=centering}
    \caption{Comparing the Performance of Spatial Pooling Methods on Compressed Images}
    \label{tab:comp_ssim_iqa_pool_results}
\begin{subtable}{0.45\linewidth}
    \centering
    \captionsetup{justification=centering}
    \subcaption{LIVE IQA (Comp) Database}
    \label{subtab:all_live_iqa_comp_results}
    \begin{tabular}{|c|c|c|c|}
        \hline
        Method & PCC & SROCC & RMSE \\
        \hline
        AM & 0.9701 &	0.9683 &	0.0699 \\
        \hline
        CoV & 0.8391 &	0.9675 &	0.1564 \\
        \hline
        \(\text{MD}^{(2,1)}\) & 0.9076 &	0.9646 &	0.1209 \\
        \hline
        FNS & 0.9675 &	0.9656 &	0.0728 \\
        \hline
        \(\mathbf{\textbf{DW}^{(1/4)}}\) & \textbf{0.9715} &	\textbf{0.9690} &	\textbf{0.0682} \\
        \hline
        \(\text{Mink}^{(2)}\) & 0.8807 &	0.9693 &	0.1364 \\
        \hline
    \end{tabular}
\end{subtable}
\hspace*{4em}%
\begin{subtable}{0.45\linewidth}
    \centering
    \captionsetup{justification=centering}
    \subcaption{TID2013 (Comp) Database}
    \label{subtab:all_tid_comp_results}
    \begin{tabular}{|c|c|c|c|}
        \hline
        Method & PCC & SROCC & RMSE \\
        \hline
        AM & 0.9438	& 0.9283	& 0.0777 \\
        \hline
        CoV & 0.9342	& 0.9505	& 0.0839 \\
        \hline
        \(\text{MD}^{(2,1)}\) & 0.9621	& 0.9499	& 0.0641 \\
        \hline
        FNS & 0.9295	& 0.9109	& 0.0868 \\
        \hline
        \(\text{DW}^{(1/4)}\) & 0.9649 & 	0.9519	& 0.0618 \\
        \hline
        \(\mathbf{\textbf{Mink}^{(2)}}\) & \textbf{0.9703}	& \textbf{0.955}	& \textbf{0.0568} \\
        \hline
    \end{tabular}
\end{subtable}
\end{table*}

\begin{table*}[t]
    \centering
    \captionsetup{justification=centering}
    \caption{Comparing the SROCC Achieved by Pooling Methods on Compressed LIVE VQA Videos}
    \label{tab:comp_ssim_vqa_pool_results}
    \begin{tabular}{|c|c|c|c|c|c|c|c|}
        \hline
        \backslashbox{TP}{SP} & AM & CoV & \(\text{MD}^{(2,3)}\) & FNS & \(\text{DW}^{(1)}\) & \(\text{Mink}^{(4)}\)\\
        \hline
        AM & 0.692	& 0.704	& \textbf{0.785}	& 0.675	& 0.694	& 0.708 \\
        \hline
        GM & 0.694	& 0.692	& 0.691	& 0.674	& 0.696	& 0.694 \\
        \hline
        HM & 0.692	& 0.680	& 0.266	& 0.616	& 0.695	& 0.688 \\
        \hline
        CoV & 0.663	& 0.139	& 0.058	& 0.650	& 0.680	& 0.004 \\
        \hline
        \(\text{MD}^{(2,1)}\) & 0.649	& 0.726	& \textbf{0.760}	& 0.636	& 0.662	& 0.678 \\
        \hline
        FNS & 0.669	& 0.677	& 0.669	& 0.655	& 0.669	& 0.677 \\
        \hline
        \(\text{W-AM}^{(99)}\) & 0.697	& 0.705	& \textbf{0.781}	& 0.684	& 0.694	& 0.708 \\
        \hline
        \(\text{W-GM}^{(89)}\) & 0.698	& 0.694	& 0.693	& 0.685	& 0.699	& 0.696 \\
        \hline
        \(\text{W-HM}^{(80)}\) & 0.699	& 0.686	& 0.660	& 0.689	& 0.699	& 0.686 \\
        \hline
        \(\text{DW}^{(2)}\) & 0.748	& 0.698	& \textbf{0.785}	& 0.723	& \textbf{0.762}	& 0.693 \\
        \hline
        \(\text{Mink}^{(1/8)}\) & 0.694	& 0.692	& 0.702	& 0.674	& 0.696	& 0.694 \\
        \hline
    \end{tabular}
\end{table*}

Knowing \(T_1^{(1)}(m,n,k)\), this sum can be computed efficiently using integral images, using the equations \eqref{eq:integral_image_start} - \eqref{eq:integral_image_end}. The temporal sum \(T_1^{(1)}(m,n,k)\) itself can be updated efficiently with each new frame, by observing that
\begin{equation}
    T_1^{(1)}(i,j,k) = T_1^{(1)}(i,j,k-1) - I_1(i,j,k-K_t) + I_1(i,j,k).
\end{equation}

In the same manner, we can also compute \(S_1^{(2)}(i,j,k)\), \(S_2^{(1)}(i,j,k)\), \(S_2^{(2)}(i,j,k)\), and \(S_{12}(i,j,k)\) efficiently. Combining these two methods, we can compute SSIM-3D in \(O(MN)\) time at each frame, irrespective of the temporal size of the window \(K_t\). 

Motion vectors were also used to incorporate temporal information in \cite{ref:mc_ssim}, which proposed a Motion Compensated SSIM (MC-SSIM) algorithm. Motion vectors were used to find matching blocks in the reference and test video sequences at each temporal index. The SSIM scores between these matched blocks were then used to calculate a temporal quality score. The clear drawback of this method is the computation of motion vectors, which are expensive and may not be readily available.

This issue was addressed in \cite{ref:three_plane_ssim}, which proposed a spatio-temporal SSIM which did not use motion information. Instead, the authors visualize the video as a 3D volume in \(x,y,\) and \(t\), where the frames lie in the \(x-y\) planes. The \(x-t\) and \(y-t\) planes contain both spatial and temporal information, and can also be compared using SSIM. The spatio-temporal SSIM (ST-SSIM) model is then defined as the average of the three SSIM values. Note that neighborhoods in the \(x-y\), \(x-t\) and \(y-t\) directions are special cases of 3-D neighborhoods used in SSIM-3D (obtained by setting the size along one dimension to 1). So, while we discuss ST-SSIM for completeness, we did not include it in our experiments. 

\subsection{Final Results}
Here, we provide comprehensive results of our experiments with the various spatial and temporal pooling algorithms described above. In all cases, we refer to each pooling method by the abbreviations listed above. To include information about the choice of optimal hyperparameters, we added superscripts to the abbreviated algorithm names. So, we denote Mean-Deviation Pooling by \(\text{MD}^{(p,o)}\), Distortion-Weighted Pooling by \(\text{DW}^{(p)}\), Minkowski Pooling by \(\text{Mink}^{(p)}\) and the Windowed AM, GM and HM algorithms by \(\text{W-AM}^{(k)}\), \(\text{W-GM}^{(k)}\) and \(\text{W-HM}^{(k)}\) respectively, where \(k\) denotes the window size.
Once again, we linearized pooled SSIM scores by fitting them with the 5PL function in \eqref{eq:five_param_logistic}, and report performance in terms of the PCC, SROCC and RMSE values.

The performances of the various spatial pooling methods on the two image databases is tabulated in Table~\ref{tab:ssim_iqa_pool_results}. On the video databases, we tested all pairs of spatial and temporal pooling methods, and these results are given in Table~\ref{tab:ssim_vqa_pool_results}. In these tables, the columns represent the choice of spatial pooling (SP) method, while the rows represent the choice of temporal pooling (TP) methods.

In Table~\ref{tab:ssim_iqa_pool_results}, the best performing spatial pooling method is boldfaced. We found that CoV pooling performed best on the challenging TID 2013 database, while the baseline Mean SSIM method performed best on the LIVE IQA database, with CoV pooling a close second.

In Table~\ref{tab:ssim_vqa_pool_results}, we boldfaced the five best results in each sub-table. It may be observed that MD Pooling and CoV pooling performed best among the spatial pooling methods, while using large windowed means performed well among the temporal pooling methods, significantly outperforming the spatio-temporal SSIM-3D and MS-SSIM-3D algorithms. However, as discussed earlier, using windowed means requires large windows (\(k \approx 50, 80\)) while providing only a minor performance improvement over the baseline. In addition, because CoV pooling performed consistently well across all databases and does not have any hyperparameters to tune, we recommend using spatial CoV pooling on picture or video frame quality maps, and the standard arithmetic mean pooling of temporal frame SSIM scores. This quality aggregation method is identical to the one used in MOVIE.

As in previous sections, we repeated the experiments on compression distorted data, and reported the results in Tables~\ref{tab:comp_ssim_iqa_pool_results} and~\ref{tab:comp_ssim_vqa_pool_results}. Even when we restricted the distortion types, we did not obtain concordant values of hyperparameters across the video databases.

Based on our recommendations, we propose a variant of SSIM, called ``Enhanced SSIM", which we are making publicly available to the community as a command line tool. The specifications of Enhanced SSIM are described below.
\begin{enumerate}
    \item Operates only on the luminance channel.
    \item Uses rectangular windows, to calculate local statistics, with a default size of 11x11. These rectangular windows are implemented using integral images.
    \item Local quality scores are computed with a stride of 5.
    \item The input image is down-sampled by a factor inferred from values of \(D/H\), using a default ratio of 3.0, corresponding to a typical \(D/H\) ratio for TV viewing.
    \item Coefficient of Variation pooling is used to spatially aggregate the local quality scores.
\end{enumerate}

We compare the performance of our implementation with LIBVMAF in Table~\ref{tab:ref_implementation_performance}, since LIBVMAF was the best performing implementation in Section~\ref{sec:ssim_versions}. This table highlights the computational and performance benefits of using our recommendations. Once again, ``(Comp)" refers to experiments conducted on compression data from each database.

\begin{table}[b]
    \centering
    \begin{tabular}{|c|c|c|}
    \hline
    Database & LIBVMAF & Enhanced \\
    \hline
    LIVE IQA & \textbf{0.9464} & 0.9377 \\
    \hline
    TID 2013 & 0.7558 & \textbf{0.7617} \\
    \hline
    LIVE VQA & 0.6954 & \textbf{0.7756} \\
    \hline
    Netflix Public & 0.7652 & \textbf{0.8360} \\
    \hline
    LIVE IQA (Comp) & 0.9543 & \textbf{0.9660} \\
    \hline
    TID 2013 (Comp) & 0.9467 & \textbf{0.9482} \\
    \hline
    LIVE VQA (Comp) & 0.6839 & \textbf{0.6936} \\
    \hline
    \end{tabular}
    \caption{Performance of Enhanced SSIM}
    \label{tab:ref_implementation_performance}
\end{table}

\section{Conclusion}
In this guide, we detailed the results of a series of experiments we conducted towards determining optimal design choices when deploying SSIM. We first evaluated the off-the-shelf performance and efficiency of several public implementations of SSIM and MS-SSIM, using which we identified a set of Pareto-optimal implementations. Using these results, we also described a method to improve the computational efficiency of SSIM using integral images. We then reviewed a method, called Scaled SSIM, to improve the efficiency of computing SSIM across resolutions, when conducting RDO in encoding pipelines. Following this, we reviewed the dependence of SSIM performance on the viewing device, where we discussed improvements to SSIM which account for viewing distance and screen size. We then investigated the dependence of SSIM on the choice of window, where we conducted extensive experiments to identify good choices for the size and type of window function (rectangular windows of size 15-20), thereby validating some observations we made when testing public SSIM implementations.

Due to the non-linear nature of SSIM, it is crucial to develop a good mapping function from SSIM scores to subjective scores. We tested a popular choice of such a mapping, the five parameter logistic function, and demonstrated its generalizability. Further, while the baseline SSIM model is defined on two luminance images, most practical applications involve media having color. To account for this, we reviewed several Color SSIM models and compared their performance, finding that Quaternion SSIM was a consistently good choice. Finally, we performed a comprehensive evaluation of spatial and temporal aggregation methods used to deploy SSIM on videos. Based on these results, we recommended using spatial CoV pooling and temporal arithmetic mean pooling of framewise SSIM scores.

In all, we have conducted a comprehensive study of many design choices involved when implementing SSIM, and made recommendations on the best practices. In addition, we have incorporated these recommendations into a variant of SSIM which we call Enhanced SSIM, for which we provide an openware command line tool for use by video quality engineers in academia and industry \href{https://github.com/utlive/enhanced_ssim}{here}.
\bibliographystyle{IEEEtran}
\bibliography{main}

\begin{IEEEbiography}[{\includegraphics[width=1in,height=1.25in,clip,keepaspectratio]{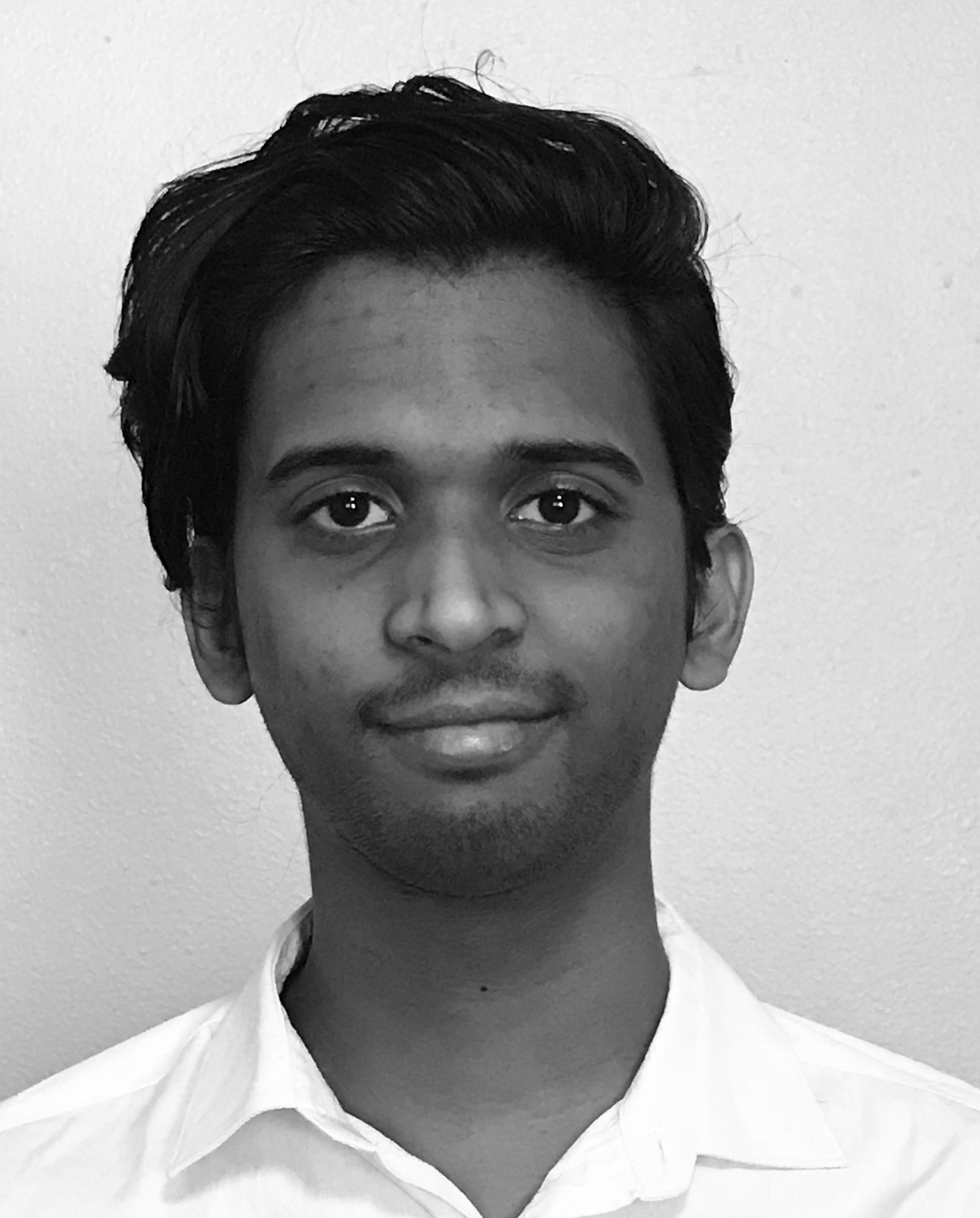}}]{Abhinau K. Venkataramanan} received his B.Tech. degree in Electrical Engineering from the Indian Institute of Technology, Hyderabad, India, in 2019. He is currently pursuing his M.S. and Ph.D. degrees in Electrical and Computer Engineering at the University of Texas at Austin, TX, USA.

During the summer of 2018, he was a summer research intern at the Robotics Institute, at Carnegie Mellon University, as an SN Bose Scholar, where he worked on biologically-inspired reinforcement learning. He is currently a Graduate Research Assistant at the Laboratory for Image and Video Engineering at the University of Texas at Austin. His research interests include image and video quality assessment, perceptual optimization, deep learning, and reinforcement learning.

Mr. Venkataramanan was a recipient of the SN Bose Scholarship (Indo-US Science and Technology Foundation) and the KVPY Fellowship (Department of Science and Technology, Government of India). He was also awarded the Institute Silver Medal by the Indian Institute of Technology, Hyderabad for securing the first rank in his department during his undergraduate studies.
\end{IEEEbiography}

\begin{IEEEbiography}[{\includegraphics[width=1in,height=1.25in,clip,keepaspectratio]{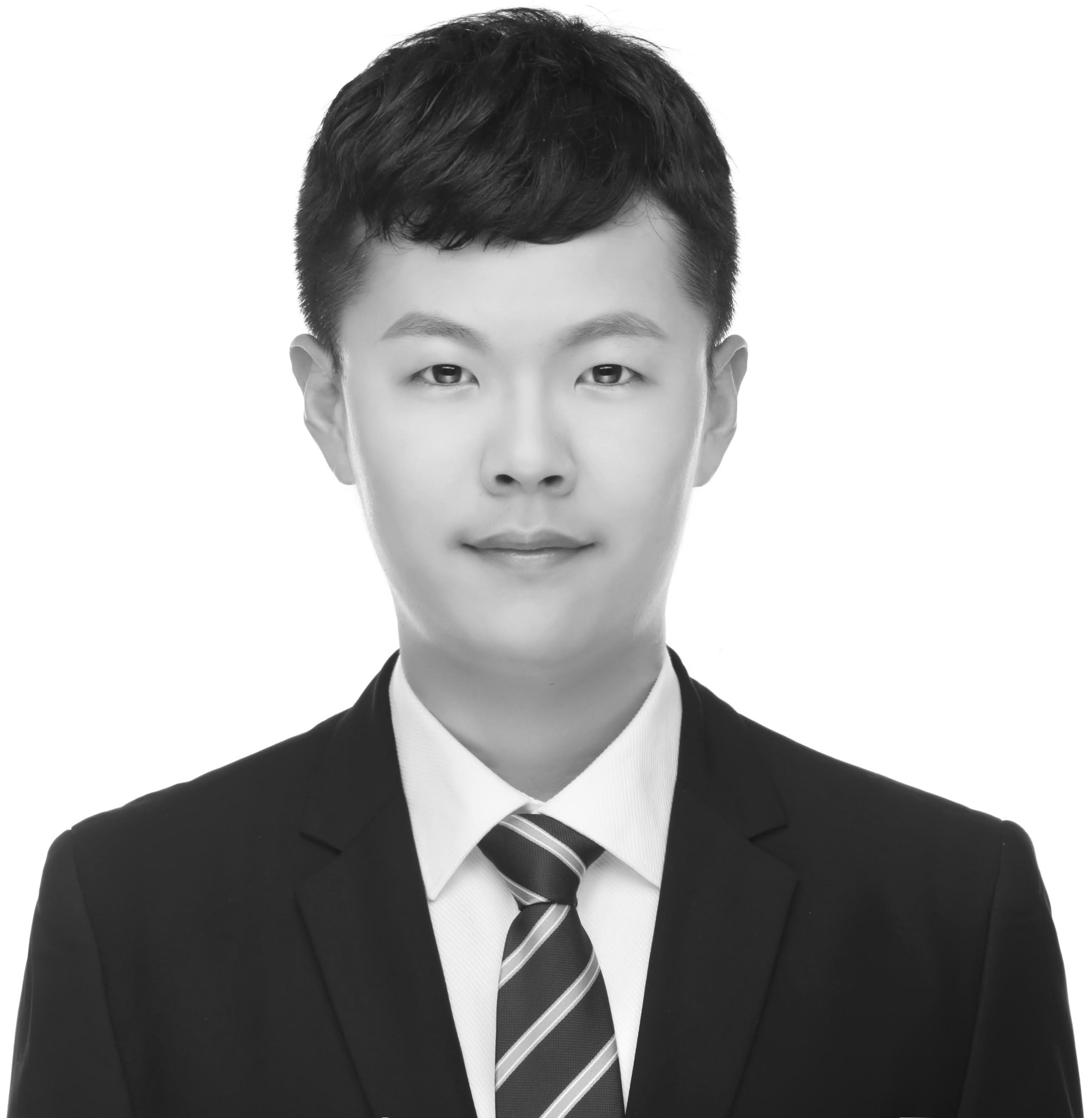}}]{Chengyang Wu} received his B. Eng. Degree in Electrical Engineering from Shanghai Jiao Tong University, in 2019. He is currently pursuing his Ph.D. degree in Electrical and Computer Engineering at the University of Texas at Austin, TX, USA.

He is currently a Graduate Research Assistant at the Laboratory for Image and Video Engineering at the University of Texas at Austin. His research interests include image and video quality assessment, no-reference methods, machine learning, and computer vision.
\end{IEEEbiography}

\begin{IEEEbiography}[{\includegraphics[width=1in,height=1.25in,clip,keepaspectratio]{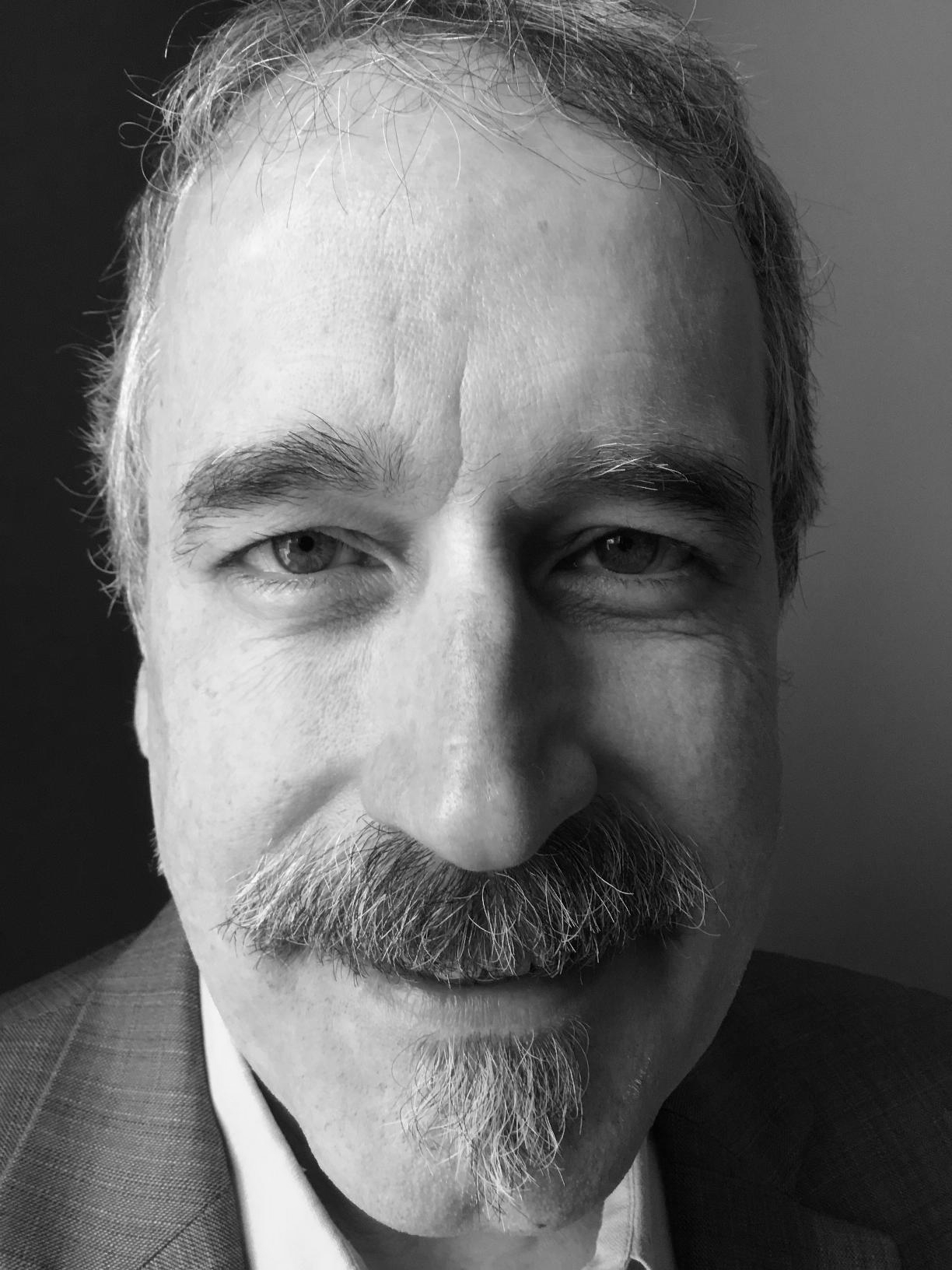}}]{Alan C. Bovik} (F ’95) is the Cockrell Family Regents Endowed Chair Professor at The University of Texas at Austin. His research interests include image processing, digital photography, digital television, digital streaming video, and visual perception. For his work in these areas he has been the recipient of the 2019 Progress Medal from The Royal Photographic Society, the 2019 IEEE Fourier Award, the 2017 Edwin H. Land Medal from The Optical Society, a 2015 Primetime Emmy Award for Outstanding Achievement in Engineering Development from the Television Academy, and the Norbert Wiener Society Award and the Karl Friedrich Gauss Education Award from the IEEE Signal Processing Society. He has also received about 10 ‘best journal paper’ awards, including the 2016 IEEE Signal Processing Society Sustained Impact Award. His books include The Essential Guides to Image and Video Processing. He co-founded and was longest-serving Editor-in-Chief of the IEEE Transactions on Image Processing, and also created/Chaired the IEEE International Conference on Image Processing which was first held in Austin, Texas, 1994.
\end{IEEEbiography}

\begin{IEEEbiography}[{\includegraphics[width=1in,height=1.25in,clip,keepaspectratio]{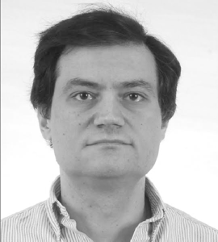}}]{Ioannis Katsavounidis} received his B.S./M.S. degree from the Aristotle University of Thessaloniki, Greece in 1991 and a Ph.D. from the University of Southern California in 1998, all in Electrical Engineering. He was an engineer for Caltech's Physics department, working on the MACRO high-energy astrophysics experiment in Italy from 1996-2000. He was a Director of Software at InterVideo in Fremont, CA, working on advanced video codecs, developing technologies around error resilience, and optimizing video encoding and decoding, from 2000-2006. In 2007, he cofounded Cidana, a mobile multimedia software company, and served as its CTO in Shanghai, China. Between 2008 and 2015, he was an associate professor at the Electrical and Computer Engineering department at the University of Thesally in Greece, where he taught and did research in signal, image, and video processing, as well as information theory. From 2015-2018, he was a Sr. Research Scientist at Netflix, part of the Encoding Technologies team, where he worked on video quality metrics and optimization problems, contributing to the development and popularization of VMAF and inventing the Dynamic Optimizer perceptual quality optimization framework. Since 2018, he is a Research Scientist at Facebook's Video Infrastructure team, working on large scale video quality and video encoding optimization problems that involve HW and SW components. He is actively involved with the Aliance for Open Media (AOM) and the next generation royalty-free codec development and the Video Quality Experts Group (VQEG) in multiple projects around video quality. Dr. Katsavounidis has over 100 publications in multiple journals and conferences and over 40 patents.
\end{IEEEbiography}

\begin{IEEEbiography}[{\includegraphics[width=1in,height=1.25in,clip,keepaspectratio]{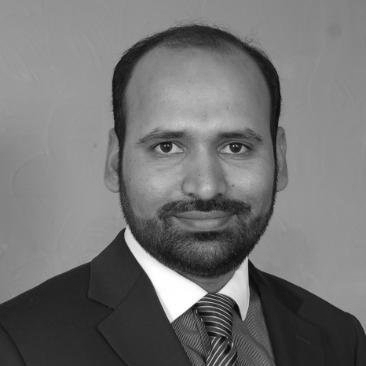}}]{Zafar Shahid} received his B.S. degree from the University of Engineering \& Technology Lahore, Pakistan in 2001, M.S. from INSA (Institut National des Sciences Appliquées) de Lyon France in 2007 and a Ph.D. from the University of Montpellier France in 2010. His PhD was funded by CNRS (Centre National de la Recherche Scientifique) France. From 2001-2006, he was a video software engineer at Streaming Networks, working on Real-time multimedia products. He was involved in video codec optimization for Philips VLIW Processor TriMedia , and developed OEM/ ODM Products for this processor. During 2011, he worked in IRCCyN Labs, Nantes France, on QOE for scalable video codecs and designed drift-free bit stream watermarking of H.264/AVC. During 2011-2015, he worked as a Media Encoding Architect for multiple start-ups designing both IPTV and ~50mSec end-to-end latency video pipelines for cloud gaming and tele-health. His video pipeline was used to stream SuperBowl 2014 to more than 500K viewers simultaneously on Desktop, iOS and Android. During 2015-2016, he worked on cloud-products for tone-mapping and metadata processing of HDR content. After this, he worked in GameStream team of Nvidia during 2016-2019. During this period, he worked on streaming of 4K video games from Cloud and from GPU enabled Gaming Desktop at home to Tegra powered Shield devices and Win/Mac clients. Since 2019, he is working at Facebook, working on video codec optimization and quality problems at scale, involving both software \& ASIC. Dr. Shahid has more than 40 publications and 3 patents. 
\end{IEEEbiography}
\EOD
\end{document}